\documentclass[a4paper,10pt]{article}
\usepackage{fullpage}
\usepackage[utf8]{inputenc}
\usepackage{authblk}
\usepackage[hidelinks]{hyperref}
\usepackage{amsmath}
\usepackage{amsfonts}
\usepackage[table]{xcolor}
\usepackage{graphicx}
\definecolor{LightGray}{gray}{0.85}
\usepackage[caption=false,font=footnotesize]{subfig}

\begin{document}

\title{Identifying user habits through data mining on call data records}

\author[1]{Filippo Maria Bianchi\thanks{filippo.m.bianchi@uit.no}}
\author[2]{Antonello Rizzi}
\author[3]{Alireza Sadeghian}
\author[4]{Corrado Moiso}

\affil[1]{\footnotesize Machine Learning Group, UiT the Arctic University of Norway, Hansine Hansens veg 18, 9019 Troms\o{} }
\affil[2]{\footnotesize Department of Information Engineering, Electronics and Telecommunications (DIET), ``Sapienza'' University of Rome, Via Eudossiana 18, 00184 Rome, Italy}
\affil[3]{\footnotesize Department of Computer Science, Ryerson University,350 Victoria Street, Toronto, ON M5B 2K3, Canada}
\affil[4]{\footnotesize Future Centre Department, in Telecom Italia, via Reiss Romoli 274, 10148 Torino, Italy.}

\date{}
\maketitle

\begin{abstract}
In this paper we propose a framework for identifying patterns and regularities in the pseudo-anonymized Call Data Records (CDR) pertaining a generic subscriber of a mobile operator. We face the challenging task of automatically deriving meaningful information from the available data, by using an unsupervised procedure of cluster analysis and without including in the model any \textit{a-priori} knowledge on the applicative context. Clusters mining results are employed  for understanding users' habits and to draw their characterizing profiles.
We propose two implementations of the data mining procedure; the first is based on a novel system for clusters and knowledge discovery called LD-ABCD, capable of retrieving clusters and, at the same time, to automatically discover for each returned cluster the most appropriate dissimilarity measure (local metric). The second approach instead is based on PROCLUS, the well-know subclustering algorithm.
The dataset under analysis contains records characterized only by few features and, consequently, we show how to generate additional fields which describe implicit information hidden in data.
Finally, we propose an effective graphical representation of the results of the data-mining procedure, which can be easily understood and employed by analysts for practical applications.
\end{abstract}

\section{Introduction}

Thanks to the popularity and wide diffusion of cellular phones, a huge quantity of mobile devices are moving everyday with their human companions, leaving tracks of theirs movements and their everyday habits. Mobile phones are becoming pervasive in both developed and developing countries and they can be a precious source of data and information, with a significant impact on research in behavioral science \cite{berry2011computational, lazer2009life}. 

A Call Data Record (CDR) is a data structure storing relevant information about a given telephonic activity involving an user of a telephonic network. A CDR usually contains spatial and temporal data and it can carry other additional useful information. Population census have been widely used in the past for keeping track of the demography and geographical movements of the population. Nowadays, due to short term and everyday mobility, more flexible methods such as various registers and indirect databases are employed: CDRs represent an optimal candidate in this sense. One of their main advantage is that they offer a statistically accurate representation of the distribution of people in an area and they can be used to track large and heterogeneous groups of people. Since CDRs evolve accordingly to the changes of users behavior, the information they carry "automatically" updates over time. 
Telecom operators continuously gather a huge quantity of CDRs, from which it is possible to extract additional information with low additional costs and generate valuable datasets. Analyses of CDR data can be successfully employed in many different fields, like monitoring the network, adaptation of supplied services (e.g., customers' billing, network planning), understanding of the economic level of a certain area, performing socioeconomic studies oriented to marketing and to build social networks \cite{duong2010constructing}.
For example, once the relationship between behavior, response, risk or other attributes is established, targeted offers of appropriate products or services can be addressed to specific customers by the telephone companies. 
Mobile positioning is a valuable source of information for investigating the spatial dynamics of human communities, but the number of published studies on this topic is still poor, mainly because of problems concerning limited access to such data and privacy issues. Localization procedures relying on mobile positioning generally provide less accurate information than the GPS (Global Positioning System), but the latter needs to be turned on to register the position, with a consequent increment of battery consumption. The wide diffusion of mobile equipment, in addition to the widespread installation of radio transmitters in both urban and rural areas, makes such positioning techniques very appealing for many case-based reasoning applications \cite{olsson2004fault}. The cellular network consists in a set of base stations formed by one tower and several directional antennae. The radio coverage of a single antenna represents a cell, whose size is not fixed in the whole network. Mobile phones can be seen as a wide-area sensor network, whose measurements can be integrated with heterogeneous sources \cite{bianchi2015prediction}. GSM (Global System for Mobile communications) is a mobile network based on the communication with an antenna covering a local area; the active connection to a certain antenna represents a spatio-temporal information which can be used for tracking the activity of a user in a GSM covered area.
An effective approach for the analysis of CDR is offered by data mining techniques based on pattern recognition and machine learning procedures, which in the last years have been successfully employed in many different fields \cite{taormina2015neural, 7286732, zhang2009multilayer, wang2015improving}.

A generic CDR relative to a telephone activity of a subscriber in a mobile network contains the identifiers of the two parties (the one which issued the call and the one which received it), the personal data of the user (name, age, sex, residence address), the coordinates of the cell which served the call, the time when the activity is registered, information on the mobile devices and on the telephonic plan. However, many of these fields are obfuscated or deleted from the publicly available records, in order to protect the privacy of the subscribers.

In fact, digital traces left by mobile phones reveal personal, often sensitive, information about their users. CDR analysis must be ruled according to the privacy of the national regulatory framework. Specifically, CDRs can be collected and processed by a mobile operator to implement the features necessary to deliver the mobile service (e.g., billing, customer care, network operation and planning). Any further usage must be either explicitly authorized by customers (e.g., through consent) or by the privacy authority. In general, the processing of anonymized CDRs (i.e., records in which the personal identification information of the referenced people is removed) is allowed, as these records are no longer personal data. Moreover, under some conditions, there is greater flexibility in processing and exploiting pseudoanonymised CDRs (i.e. records in which the ID of the referenced people is replaced with a code, often obtained through one-way cryptography): in particular, the pseudoanymization procedure must prevent the re-identification of a person from the analysis of the pseudoanonymized records. These conditions set some constraints on the datasets that a mobile operator can analyze or transfer to third parties. Therefore, also in case of pseudoanonymized records, the dataset must be pre-processed in order to reduce probability of re-identification. A common procedure consists in decreasing time resolution or increasing space granularity, so that the data collection never spans long time periods or the spatial information is not detailed. Typical examples are datasets of CDRs with high spatial resolution containing records of users which are monitored for a short period of time, or datasets where user activities tracked for longer time intervals usually come with a lower space resolution. This latter is the case considered for this study and it will be discussed in details in Sect. \ref{sec:dataset_desc}.

In this paper we propose a data-mining procedure for automatically identifying the recurrent patterns in the telephonic activity of mobile network users, in order to understand and describe their habits. We design an inference system that uses a cluster-based approach to discover regularities among data. Cluster analysis can be framed into unsupervised learning, which is the task of identify hidden structures in unlabeled data. As in our case of study, the ground truth of the expected result is unknown and there is no error or reward signal to evaluate a potential solution.
Cluster-based approaches have been successfully applied for the discovery of new concepts in streams of data \cite{Spinosa:2007:OCA:1244002.1244107, japkowicz1999concept}. However, the outcome of the clustering procedure is strongly influenced by the dissimilarity measure adopted and, in general, it depends on a set of configuration parameters. The procedure of tuning such parameters could be difficult and it may requires \textit{a-priori} information not always available.
As the core inference engine for our application, we use the LD-ABCD algorithm, which has been recently developed by the authors of this paper. LD-ABCD implements a novel cluster-analysis procedure, which has been presented in \cite{bianchi2014agent}. In order to validate the effectiveness of the proposed approach, the algorithm has previously been tested on synthetic and benchmarking datasets, where the ground-truth was known. In this work, we integrate our newly-designed system into a larger framework, which has been specifically designed to deal with a novel real-world case of study. Since we do not possess a supervised information on the results, we evaluate the performances of our system through a comparison with a well-established subspace clustering algorithm.
The main contributions of this work are summarized in the following.

\begin{enumerate}
		\item We propose a cluster-based approach for retrieving multiple groups of CDRs, which are similar according to different subsets of features. We do not make assumptions in advance on which characteristics should be taken into account for identifying clusters, or on the total number of clusters. Moreover, each cluster is characterized by its own dissimilarity measure parameters, according to the concept of \textit{local metric learning}.
	We interpret the well-defined clusters that have been identified as relevant patterns in the activity of a given user. Such patterns are used to generate a \textit{digital fingerprint} representing user's habits, in terms of telephonic activity, geographical movements, time periods when daily communication activities are more frequent, most visited places (home, workplace etc..) and a social profiling. These fingerprints can be employed for different purposes, like profiling and definition of classes of users, depending on the specific application. With respect to other works focused on the analysis of CDR, we propose a new framework based on a complex data mining and knowledge discovery procedure. We show how meaningful patterns can be extracted and used to characterize a user, preserving his privacy and without making any \textit{a-priori} assumption on the nature of the data. 
	\item When data characterized by an high number of distinct features are available, many informative, significant and useful information can be easily derived,	more complex analysis can be performed and non-trivial relationships among data can be discovered. 
However, in our work we process a dataset of pseudo-anonymized CDRs where each entry contains only a limited number of attributes. The problem we face is challenging since it seems that, at a first glance, only naive regularities in the data can be retrieved.
In this paper we show how to extract implicit information from the data and we use them for identifying hidden frequent patterns which lead to meaningful results and considerations.
		\item We propose an effective method of visualization, which encodes data and information into visual objects. Our main goal is to communicate information clearly and effectively through graphical tools, in order to express and to quantify the results , through visual human interfaces \cite{fayyad2002information}.
\end{enumerate} 

The remainder of the paper is organized as follows: in Sect. \ref{sec:otherworks} we review some relevant works and applications focused on the analysis of CDRs. In Sect, \ref{sec:dataset_desc} we present the dataset considered for the analysis, discussing the representation of the data and how implicit information contained in the CDRs can be extracted. In Sect. \ref{sec:method} we propose a framework that can be used for discovering relevant patterns in the data by relying on a procedure of cluster analysis and we show the obtained results in Sect. \ref{sect:experiments}. Finally, in Sect \ref{sec:conclusion} we draw our conclusion and we discuss future works and improvements. 

\section{Works on Call Data Records}
\label{sec:otherworks}

In this section we review some relevant works which leverage the information contained in CDRs for a multitude of different applications.

CDRs can be effectively used for understanding the interactions between users and to define a network of social relationships among them. In \cite{eagle2009inferring} the authors use CDRs information to provide insights into the relational dynamics of individuals, demonstrating that it is possible to accurately infer friendships. They proved that calls between friend dyads have distinctive temporal and spatial patterns. In \cite{candia2008uncovering}, the authors use CDRs to study the average collective behavior at large scales, focusing on the occurrence of anomalous events.

One of the largest field of application of CDRs is the prediction of the movements of the users according to their calls. By predicting their movements, is it possible to understand the life-style of the users and to provide useful information to the telephonic companies, in order to tailor targeted phone plans for the customers. CDRs are used for understanding which are the most busy areas and times during the day, to predict the movements of the people and promote a better transportation service \cite{berlingerio2013allaboard}. In \cite{calabrese2011real} mobile data are used for monitoring in real-time the traffic conditions and pedestrian distribution. In a study performed by the German telephone company T-Mobile, mobile data are used for tracking trajectories of vehicles on the streets \cite{schlaich2010generating}. A classification of the users based on their movements is proposed in \cite{furletti2013analysis}, where the customers are assigned to three predefined classes (resident, commuters and visitors), depending on their movements in certain areas. The raw CDRs are initially aggregated on both spatial and temporal basis: for each user the authors build a matrix containing information on the number of calls which took place on a given area in three different periods of the day. Raw data can be aggregated directly by the telephone operator, in order to generate a compact representation which is easy to manage and above all is anonymized, so that the privacy of the users can be preserved. On the aggregated data they perform a clustering procedure using Self Organizing Maps and the obtained clusters are labeled with the most frequent class. Such clusters can be used to describe the habits of the users in a given area.
In \cite{song2010limits} the limits on the predictability of the movements of the people are studied. The authors used CDRs gathered from 50.000 users in a time-window of 3 months of activity and they propose three different measurements of entropy for analyzing the predictability of their movements: the random entropy, the temporal-uncorrelated entropy and the actual entropy. They claim to be able to predict 93\% of the users movements at best and 80\% in the worst case.
In \cite{gidofalvi2012and}, authors build an Inhomogeneous Continuous-Time Markov (ICTM) model using both temporal and spatial information, for predicting the next position of the user in a given time lapse. The results show that the ICTM model is able to forecast the correct time interval with an error of 45 minutes and the next location with a 67\% accuracy.
Also in \cite{citeulike:11530535} the authors suggest using spatio-temporal data for predicting users position, employing 10 different types of models based on Bayesian rules or Markov models. The best forecast accuracy is achieved with a model called HPY (Hierarchical Pitman-Yor) Prior Hour-Day Model, which is able to correctly localize the users with an accuracy of 50\%.
In \cite{ozer2014predicting} the urban cells are clustered together, in order to avoid  repeated switches of users between adjacent cells and to force them to assume the same size of the rural cells. Successively, a data mining procedure is performed on the sequences of the movements, aiming to predict whether a user will change its position or not. In order to increase the accuracy, temporal information are also included. The CDRs are then represented with one-day-long sequences and 3 different problems are considered: (i) predicting the first time interval when the user will change his position and where he will move; (ii) predicting only the next movement; (iii) predicting the next movement and his next telephonic activity.

Other applications leverage CDRs to promote analysis, diagnosis and prevention in organizational, social and security contexts, like tracking population movements in emergency conditions to improve government alert communications. Social applications of the CDRs involve statistics on the population, performed through a correlation of mobile data with poverty indexes and GNP (Gross National Product) in order to draw poverty maps that can be used for orienting actions and programs for development in the most needing regions \cite{smith2013ubiquitous}. Correlating mobile data with information on the diffusion of epidemics and diseases, allows the definition of efficient models for preventing and containing the spreading of epidemics. Such models can be employed for informing population, optimizing medical resources and planning vaccinations \cite{lima2013exploiting}.

CDRs can also be used for profiling or for understanding the behaviors of one or more users. An interesting application of CDRs is the analysis and prediction of the lifestyle of the users. In these studies, the privacy of the users must be protected, their identities must remain unknown and not be accessible from outside. Those aspects are treated in \cite{StefvandenElzenD4D}, where it is proposed a strategy for increasing the security of the data, preserving data usability and their semantic value. In \cite{grindrod2013infering} the authors propose a model for grouping users according to their similarity in calling patterns and for classifying new call records. The procedure consists in (1) extracting from the CDRs the attributes which better characterize each user; (2) applying a clustering algorithm to identify common behavioral patterns among the costumers, trying to balance the dimension of clusters, each one representing a different state; (3) generating a transition matrix among the states which must be sparse, so that a user can reach only a limited subset of different states each time; (4) using such model to classify new call records. The model reflects the habits of the user and it is uncorrelated with his identity, meaning that it can be exported and re-used. The model can also be dynamically updated as long as new data are collected, evolving as the life-style of the user changes. Such model can be used by telephone companies for offering products to the user, according to their current and potential future states.
In \cite{hu2013we} the authors classify the users in a set of pre-determined classes by relying on the CDRs. Only the fields of the CDR which discriminate better the users are selected. In \cite{ahas2010using}, CDRs are used to identify the areas most visited by the users, called "anchor-points" (AP), which are commonly the working and the living place along with the locations where secondary activities take place with a given frequency (e.g. the gym). The procedure for identify AP is the following: (1) for each user they identify the cells from where more than 2 calls are issued on a monthly basis (Regular Cells - RC); (2) the two RC with the highest number of calls are tagged as home and work place. Work place is discerned from home place depending on the timing of the calls; (3) they process cases of users with more than one work or home AP, which could live or work on borders of adjacent cells; (4) for users whose activity is mainly registered in a single cell -- meaning, for example, that they live and work in the same area -- a multi-functional task AP is defined; (5) the remaining cells are classified as secondary AP.


\section{Dataset Description}
\label{sec:dataset_desc}
In this work we analyze a dataset of the Orange telephone data published for the "Data for Development" (D4D) challenge \cite{DBLP:journals/corr/abs-1210-0137}, which is an open collection of CDRs, containing anonymous calling events of Orange's mobile phone users in Ivory Coast. More information on the challenge are available on the website \url{http://www.d4d.orange.com}. The data consist in anonymized mobile phone calls and SMS that have been gathered in the period that spans form December 1, 2011 to April 28, 2012 and they are arranged in four different datasets. We analyze the CDRs relative to individual trajectories of 50,000 randomly selected customers, gathered over the entire observation period. Each CDR contains the time of the call and the location, expressed as the identifier of the prefecture, from where the call has been issued. Even if the spatial resolution in the records is low, the users are tracked for a time interval which is sufficiently long to identify meaningful patterns in their activity. This aspect is fundamental, since analyzing data concerning the activities of a user for an extended time interval allows a better detection of the regularities in his behavior and the profile can be drawn with higher accuracy. 

A CDR in this dataset has the following structure: \{\texttt{user\_id}, \texttt{conn\_datetime}, \texttt{subref\_id}\}, where \texttt{user\_id} is the anonymized identifier of the user, \texttt{conn\_datetime} is the date and the time of the registered telephonic activity and \texttt{subref\_id} is the identifier of one of the 255 sub-prefectures in the country.

\subsection{Data preprocessing and representation}
\label{sec:datarep}

The data that we consider contain only two fields, which are the temporal and the spatial information relative to each registered telephonic activity. However, it is possible to extract useful implicit information contained in the dataset, adding additional features. The the resulting final structure of an element in the dataset is: \{ \texttt{subref\_id}, \texttt{week\_day}, \texttt{work\_day}, \texttt{conn\_time}, \texttt{day\_period}, \texttt{prev\_call} \}. In the following, we provide a description for each field in the augmented records.

\begin{itemize}
\item \texttt{subref\_id} is preserved as in the original dataset. Since the values are unique identifiers, we consider the domain for this field to be discrete (nominal).
\item \texttt{week\_day} is generated from the field \texttt{conn\_datetime} and it represents the day of the week where the connection has been registered. The possible values are: \{ \texttt{Mon}, \texttt{Tue}, \texttt{Wed}, \texttt{Thu}, \texttt{Fri}, \texttt{Sat}, \texttt{Sun} \}. They are drawn from a nominal domain.
\item \texttt{work\_day} is a Boolean value derived from \texttt{conn\_datetime} and it distinguishes working days from weekend days (0 for Saturday and Sunday, 1 for the other days). Also in this case the domain is nominal.
\item \texttt{conn\_time} is the time, expressed in hours and minutes (\texttt{HH:MM}), when the telephonic activity is registered. This is a continuous circular domain in [\texttt{00:00}, \texttt{23:00}] with resolution of 1 hour.
\item \texttt{day\_period} is the period of the day when the call was issued. In particular, activities registered in the time interval [\texttt{07:00}, \texttt{13:00}] are considered morning activities, the ones in [\texttt{14:00}, \texttt{19:00}] are afternoon activities and the ones in [\texttt{20:00}, \texttt{06:00}] are evening/night activities. We have then a nominal domain with the values \{ \texttt{Mor}, \texttt{Aft}, \texttt{Eve} \}.
\item \texttt{prev\_call} is the elapsed time (in minutes) from the previous call and it belongs to an ordinal domain in $\mathbb{N}^+$.
\end{itemize}

We represent the CDR with a data structure called \emph{sectioned vector} \cite{komorowski1997principles}, which can be defined as a \textit{n}-dimensional vector $\textbf{u}$ composed of the concatenation of $s$ opportune sub-vectors, called \textit{sections}. The sectioned vector is a flexible data structure, which allows to represent an object whose components belong to different domains, tailoring the definition of a proper dissimilarity measure for each section. The number and size of the sections constitute the structure of the sectioned vector:
\begin{equation}
\textbf{u}=(\textbf{u}_0; \textbf{u}_1; \ldots; \textbf{u}_{s-1}),\ \mbox{with}\ \sum\limits_{i = 0}^{s - 1} {\dim (\textbf{u}_i)}  = n.
\end{equation}

Other existing alternatives to represent data containing values coming from heterogeneous domains are labeled sequences and graphs. These data structure also carry topological information on data, but they are more difficult to handle and the procedures used for evaluating their dissimilarity are usually complex, less accurate and more demanding in terms of computational resources \cite{bianchi2013matching, bianchi2014granular}. 
In many cases data are not characterized by important topological information, i.e. it is only relevant the numerical value of each component and not their spatial/temporal organization. This coincides with our case of study, for which it is recommended to use sectioned vectors, avoiding more complex data structure. 

We now introduce a generalized dissimilarity measure for sectioned vectors of $\mathbb{R}^n$. An highly flexible dissimilarity measure is required when the components have heterogeneous meaning and the use of a simple Minkowski metric is not so obvious \cite{li2002dpf,6399461}. Let $\textbf{u}^{(a)}, \textbf{u}^{(b)}$ be two sectioned vectors with the same structure, and let $d_i(\textbf{u}_i^{(a)}, \textbf{u}_i^{(b)})$ be a dissimilarity measure for the $i$-th section. We can define the dissimilarity measure for the whole vectors $\textbf{u}^{(a)}$ and $\textbf{u}^{(b)}$  as: 
\begin{equation}
\label{eq:diss}
d(\textbf{u}^{(a)} ,\textbf{u}^{(b)}, \mathbf{m} ) = \sum\limits_{i = 0}^{s - 1} {\mathbf{m}(i) d_i (\textbf{u}_i^{(a)} ,\textbf{u}_i^{(b)} ),} 
\end{equation}
being $\mathbf{m}$, in general, a real-typed vector in the unitary hypercube in $\mathbb{R}^s$, which represents the parameter configuration (PC) of the dissimilarity measure $d(\cdot,\cdot,\mathbf{m})$. Each entry is a weight that tunes the importance of the respective component in the structured vectors for the computation of the total dissimilarity. For example, if $\mathbf{m}(1)$ assumes an high value (close to 1) it means that the difference between the values in the first section of two vectors $\textbf{u}_i^{(a)}$ and $\textbf{u}_i^{(b)}$ highly influences the degree of their total dissimilarity. 

In our dataset, the content of the sections belong to three types of domain: ordinal, nominal and ordinal circular. Consequently, we use three different dissimilarity functions, which are introduced in the following.
For what concerns the ordinal domain from which the values of \texttt{prev\_call} are drawn, we use the normalized Manhattan distance.
The decision of using the Manhattan distance was maturated after having tried different type of dissimilarity measures, among which the Euclidean distance. In our preliminary experiments we observed that there were not significant improvement (or in general changes) in the quality of the results obtained. On the other hand, there was a tangible increment in the computational time. The evaluation of the dissimilarity measure is repeated many times during the execution of the algorithm, which benefits from the simplicity of the Manhattan distance, which requires performing only sums and multiplications with respect to exponential operations.
The Manhattan distance is defined as follows:

\begin{equation}
\label{eq:manhattan}
d_M (\textbf{u}_i^{(a)} ,\textbf{u}_i^{(b)} ) = \frac{\sum\limits_{j = 0}^{\dim (\textbf{u}_i) - 1} {\left| {u_{ij}^{(a)}  - u_{ij}^{(b)} } \right|}}{\max{(\textbf{u}_i)} - \min{(\textbf{u}_i})}.
\end{equation}

For comparing the values of \texttt{subref\_id}, \texttt{week\_day}, \texttt{work\_day} and \texttt{day\_period} that belongs to a nominal domain, we use the Delta dissimilarity function, defined as:

\begin{equation}
\label{eq:delta}
d_{\Delta} (\textbf{u}_i^{(a)} ,\textbf{u}_i^{(b)} ) = 
\begin{cases}
0 & \mathrm{if}\ \textbf{u}_i^{(a)} = \textbf{u}_i^{(b)}, \\ 
1 & \rm{otherwise.} \\
\end{cases}
\end{equation}

Finally, for what concerns the values of the attribute \texttt{conn\_time}, it is required a dissimilarity that operates on an ordinal circular domain. Specifically, we defined the following function:

\begin{align}
\label{eq:circ}
\begin{split}
& d_{C} (\textbf{u}_i^{(a)} ,\textbf{u}_i^{(b)} ) = \frac{\min{ (t_1, t_2 )}}{12}, \text{with}\\
&t_1 = \| \textbf{u}_i^{(a)} - \textbf{u}_i^{(b)} \|,\\
&t2 = \min{ (\| 24 - \textbf{u}_i^{(b)} + \textbf{u}_i^{(a)} \|, \| 24 -\textbf{u}_i^{(a)} + \textbf{u}_i^{(b)} \|) }.
\end{split}
\end{align}

In order to stress the importance of a section in the computation of the dissimilarity measure, in this work we use PCs with Boolean weights, i.e. $\mathbf{m} \in {0,1}^6$. Even if defining $\mathbf{m}$ in the continuous interval [0,1] provides an higher degree of flexibility, we opted for Boolean weights to facilitate the interpretation of the features which are selected to be relevant. 

Defining $\mathbf{m}$ in advance is a difficult task which demands \textit{a-priori} information on the data and on the applicative context. If, as in our case, prior knowledge is not available and the objective is to identify \textit{any} existing regularity in the data, it is required a strategy for effectively determining the value of weights. This can be done through a data mining procedure, which deals with the problem of metric learning \cite{6588959,4674367,yin2012semi,zhang2012semi,chang2012boosting} and the concept of local metrics \cite{kim_duin__multiplediss__2009,6399461,Queiroz20133383,Bereta20131213,gonen2011multiple}. 

The focus of our work is to determine multiple instances of $\mathbf{m}$, which allow to retrieve meaningful regularities among data, necessary to build \textit{digital fingerprints} of the users. 
We use a data mining algorithm that has been designed to automatically determine a specific dissimilarity measure for each cluster, implementing a \textit{local} distance learning approach.


\section{Approaches for the analysis}
\label{sec:method}

In this section, we present a framework that relies on two different cluster-based approaches \cite{jain1988algorithms, yager1994generation, demirli2000higher} to retrieve relevant information in the dataset of CDRs.

When data are represented by structured vectors and the PC $\mathbf{m}$ of the dissimilarity measure is the collection of Boolean weights that determine the importance of each feature in the evaluation of the similarity of two elements, the task of setting $\mathbf{m}$ can be seen as the selection of the relevant features in each cluster. In this case, tuning the values in $\mathbf{m}$ can be straightforwardly connected to the problem of \textit{subspace clustering}, which retrieves the subset of dimensions to be considered in the clustering problem, by removing irrelevant and redundant dimensions. 
While traditional clustering algorithms consider all the dimensions in the attempt of collecting as much information as possible, in high dimensional data many of the dimensions are often irrelevant. Irrelevant dimensions can misguide the clustering procedures by hiding clusters in noisy data. Unlike feature selection methods which examine the dataset as a whole, both the approaches that we present in this section localize their search and are able to uncover clusters that exist in multiple, possibly overlapping subspaces (local metrics learning). Clusters found in lower dimensional spaces also tend to be more easily interpretable, allowing the user to better conceive further analysis.

This procedure is also highly related with feature selection and dimensionality reduction techniques \cite{zhang2009dimension}; it can be considered as an extension that attempts to find clusters in different subspaces of the same dataset. 
Just like the feature selection procedure, the algorithms that we present require a search method and an evaluation criteria and they must somehow limit the scope of the search, to consider different subspaces for each cluster.

We used a data visualization technique, based on charts of various types, to show patterns and relationships in the data for one or more variables. Through an effective visualization, users can easily analyze data, discovering interesting insights. The visualization method that we propose in Sect. \ref{sect:experiments} makes complex data more accessible, understandable and usable by a larger class of users, not necessarily experts in data mining and graphics reading. 

The first algorithm that we discuss in Sect. \ref{sec:ldabcd_desc}, is a novel multi-agent system for cluster and knowledge discovery called LD-ABCD \cite{bianchi2014agent}, which has been recently developed by the authors of this paper. LD-ABCD identify well-defined clusters in the dataset using different configurations of the dissimilarity measure and it provides a semantic characterization of each identified cluster.
The second method that we consider as an alternative implementation of our framework, is PROCLUS, discussed in Sect. \ref{sec:proclus_desc}, a subspace clustering method that identifies the most descriptive dimensions among the elements in each cluster.

\subsection{Cluster Discovery by Dissimilarity Function Adaptation}
\label{sec:ldabcd_desc}

LD-ABCD is a multi-agent algorithm designed to automatically discover relevant regularities in a given dataset, determining at the same time a set $\mathcal{M}$ of PCs of the adopted parametric dissimilarity measure, yielding compact and separated clusters in the data.
Each agent operates independently on a suitable weighted graph, which is used to represent the data. The graph is fully connected, each node corresponds to an element in the dataset and the edges are labeled with a value proportional to the similarity of the two connected elements. 
The weights on the graph depend on the specific PC $\mathbf{m}_j$ of the dissimilarity measure adopted by the $j$-th agent. A new PC is iteratively selected by an agent $j$ to construct its own instance $G_j$ of the weighted graph.

The clusters discovery procedure is implemented by means of multiple Markovian random walks (RW), which are performed independently, at the same time, by several agents on the different instances of the weighted graph.
The behavior of the RW is thus dependent on the PC selected by the agent. During a RW, an agent searches and takes decisions autonomously for one cluster at a time. 
A suitable on-line mechanism is designed to decide whether a set of elements visited (i.e., "walked upon") by an agent should be accepted as a meaningful cluster or rejected. Specifically, the set of nodes visited by the agent during the random walk is a subgraph $c_i$ that represent a potential cluster. Its quality is evaluated with a measures called Cluster Quality (CQ) that relies on the concept of the graph conductance \cite{kannan2004clusterings} and that depends also from the configuration $\mathbf{m}_j$ of the dissimilarity measure currently adopted by an agent, such that CQ is a function of the pair $\langle c_i, \mathbf{m}_j \rangle$. A cluster is accepted by an agent if its CQ value is greater than a threshold $\tau_{\text{CQ}}$, which controls the overall quality of the solutions returned by the algorithm.

An edge $e_{kl}\in\mathcal{E}$ is labeled with a weight, $w(e_{kl}; \mathbf{m}_j) \in [0, 1]$, which depends on the dissimilarity $d(x_k, x_l; \mathbf{m}_j)$ according to the following relationship:
\begin{equation}
\label{eq:edgelabel}
w(e_{kl}; \mathbf{m}_j) = \text{exp}(d(x_k, x_l; \mathbf{m}_j) \cdot \tau_{\text{exp}})
\end{equation}
where $\tau_{\text{exp}}$ is a parameter used to magnify the edge weights between similar elements, making less likely the unwanted transitions to vertices connected by low weights. A correct setting of $\tau_{\text{exp}}$ is crucial, since it affects the behavior of the RW. An heuristic approach proposed in \cite{bianchi2014agent}, consists in generating a weighted graph for every value of $\tau_{\text{exp}}$ in an interval $[\tau_{\text{exp}}^{min}$, $\tau_{\text{exp}}^{max}]$ and then performing a sufficiently high number of RWs to retrieve clusters. The optimal $\tau_{\text{exp}}$ is evaluated in function of the average CQ and cardinality of the clusters returned.

As long as the execution of LD-ABCD proceeds, an agent might find a set $\mathcal{C}$ of very similar (or even equal) clusters using different PCs, in the sense that they may overlap significantly. These clusters are merged into a \emph{meta-cluster} $\hat{c}$, which contains the elements which appear more frequently in the clusters in $\mathcal{C}$. All the PCs used to discover the clusters in $\mathcal{C}$ are considered \textit{equivalent} with respect to $\hat{c}$ and are inserted in a list $\mathcal{L}$ associated to the meta-cluster. 
The CQ value of each PC $\mathbf{m}_j \in \mathcal{L}$ is then recomputed on $\hat{c}$ and the higher its value, the better $\mathbf{m}_j$ characterizes the elements in $\hat{c}$. A parameter $\vartheta$ is used as a threshold for the distance between two clusters, in terms of percentage of elements commonly shared, for defining whether two clusters should be considered similar and included in the same meta-cluster; notably, $\vartheta$ can be thought as the "radius" of the meta-cluster.

The final output of the algorithm is a collection of meta-clusters, which may be (partially) overlapped and which may not cover the entire dataset. Each meta-cluster is associated with the list of PCs $\mathcal{L}$, representing a consistent and interpretable semantic characterization of the meta-cluster, since it defines the set of dissimilarity measures according to which the elements in the meta-cluster as indistinguishable.
An important remark is that we forced each PC to assume at least two values different from 0, in order to avoid trivial solutions containing clusters composed of elements which are similar only with respect to a single feature. 

Thanks to the multi-agent architecture, the algorithm possess high scalability and a distributed implementation is straightforward. LD-ABCD can process any type of data (vectors, graphs, sequences, etc..), given a suitable (parametric) dissimilarity measure for comparing the elements. In particular, the algorithm can process data that are defined in non-metric spaces, greatly increasing the scope of problems that can be treated. 
Unlike the $k$-clustering paradigm, LD-ABCD does not require to specify in advance the desired number of clusters to be returned. This feature is very important in our case of study, since we do not possess any \textit{a-priori} information on the dataset.
Another significant feature of LD-ABCD, which is relevant for our application, is the provided powerful semantic characterization of the identified clusters, allowing important analyses on the content of the results returned.

\subsection{Subspace Clustering}
\label{sec:proclus_desc}

The alternative procedure that we consider for comparing performances of our framework is based on a subspace clustering algorithm. Subspace clustering algorithms can be divided in two groups, the top-down search and bottom-up search methods, which are distinguished by their approach to identify subspaces \cite{parsons2004subspace}. 

The bottom-up search methods try to reduce the search space, taking advantage of the downward closure property of density: if there are dense units in $k$ dimensions, there are also dense units in all $k-1$ dimensional projections. These methods first create an histogram for each dimension and then select those bins whose densities are above a given threshold. Candidate subspaces in higher dimensions can then be formed using only those dimensions that contain dense units, dramatically reducing the search space. The algorithm proceeds until no more dense units are found. Adjacent dense units are then combined to form clusters. These algorithms, of which CLIQUE \cite{agrawal1998automatic} is one of the most famous representative, have been conceived to be applied on strictly real-typed domains, where a notion of proximity can be easily defined. In our case, we use sectioned vectors containing values drawn from circular and discrete domains, where the concept of density cannot be defined. 
This prevents to map similar values of the domain to close position on a grid representation. For this reason, applying bottom-up search methods to our problem is not feasible and we decided to consider the top-down approach.

The top-down subspace clustering approaches start by finding an initial approximation of the clusters in the full feature space with equally weighted dimensions. Successively, to each dimension is assigned a weight for each cluster. The updated weights are then used in the next iteration to regenerate the clusters. This approach requires to repeat multiple iterations of the clustering algorithm, considering the full set of dimensions, in order to converge to the optimal solution. 

Among all the top-down approaches, we selected the PROCLUS algorithm \cite{Aggarwal:1999:FAP:304181.304188}, which is one of the first developed and probably the most famous. The algorithm consists in three different steps called initialization, iteration, and cluster refinement, during which the clustering is iteratively improved. 
In the initialization step PROCLUS samples the data, then it selects with a greedy strategy a set $\mathcal{M}$ of representatives, called medoids, to be as much spread as possible in the full dimensional space. The medoids represent the pool of the candidates for the cluster representatives. 
In the iteration phase, a random set of $k$ medoids are selected from $\mathcal{P}$ and for each medoid $p_i$ a neighborhood is generated, consisting in all the points whose distance from $p_i$ is less than or equal to the distance from $p_j$, being $p_j$ the closest medoid to $p_i$. For each medoid, the algorithm selects the set of dimensions along which the distances of the elements in the neighborhood are the smallest. The total number of dimensions associated to medoids must be $k \cdot l$, where $l$ is an input parameter that selects the average dimensionality of the subspaces for each cluster. Once the subspaces have been selected for each medoid, all the elements in the dataset are assigned to their closest medoid, according to the average Manhattan segmental distance, which considers in the computation of the dissimilarity from each medoid only the selected dimensions.
The medoid of the cluster with the least number of points is discarded along with any medoids associated with fewer than $(\frac{N}{k}) \cdot \texttt{minDev}$ points, being $N$ the total number of patterns and \texttt{minDev} another input parameter. At the end of each iteration, discarded medoids are replaced with new ones, randomly chosen from $\mathcal{P}$, and it is checked if clustering has improved.
In the refinement phase, PROCLUS computes new dimensions for each medoid based on the clusters formed (rather than on the neighborhood) and then it reassigns points to the medoids, removing outliers. 

Due to the use of sampling, PROCLUS is faster than many other subspace clustering algorithms, especially on larger datasets. On the other hand, one of the main drawback of the algorithm is its strong dependence on the parameters $k$ and $l$ which, in many cases, can be hard to be set in advance, since they require an adequate knowledge of the problem and of the dataset at hand. Another drawback is due to the bias toward clusters that are hyper-spherical in shape. Additionally, since the average number of dimensions is given, the number of selected dimension in each cluster will be similar. It is also important to notice that PROCLUS creates a partition of the dataset and, possibly, an additional group of outliers. This means that each instance is assigned to only one cluster (or to the outliers group). This is a critical difference with respect to the procedure implemented in LD-ABCD, which does not form a proper partition, allowing the generation of overlapping clusters, meaning that an element can be assigned to one cluster, more cluster or no clusters at all.

In our experiments we used PROCLUS configured with the dissimilarity measure defined in Eq. \ref{eq:diss}, which can be considered a generalization of the average Manhattan segmental distance.

\section{Experiments and Results}
\label{sect:experiments}

In this section, we first analyze the set of CDRs relative to a specific user. Successively, we process the dataset with our data-mining procedure and we discuss the results obtained by the two different implementations, proposed in Sect. \ref{sec:method}.

The original dataset presented in Sect. \ref{sec:dataset_desc} contains CDRs relative to 50,000 users. In our experiments we processed the data relative to the calls of more than 100 different users, which have been randomly selected. For most of them it was possible to identify clear and distinct patterns, while for others we did not obtained meaningful results, mainly because of irregular telephonic activity or for the limited number of the calls issued by the user. In the following, we show an example of how an analysis of the CDRs relative to a given user can be performed using the proposed methodology.

To visualize the content of the CDRs relative to a given user, we use 4 different charts that show how the CDRs are distributed according to the accounted features. In particular we have:

\begin{enumerate}
	\item An histogram describing the number of calls done by the user from different prefectures. Each bin is associated with one of the prefectures from where calls were issued and the height of the bars is proportional to the number of calls done in that prefecture.
	\item An histogram that describes the distribution of the values contained in the field \texttt{prev\_call}. Each bin of the histogram represents the time elapsed from the previous call and its height is proportional to the number of CDRs whose value \texttt{prev\_call} falls in that interval. 
	\item An histogram that represents the distribution of the calls of the user among the 7 days of the week, according to the field \texttt{week\_day}.
	\item An histogram that represents the distribution of the calls of the user among the 3 periods of the day, according to the field \texttt{day\_period}.
\end{enumerate}

\begin{figure*}[!ht]
\centering
\includegraphics[width=0.6\textwidth]{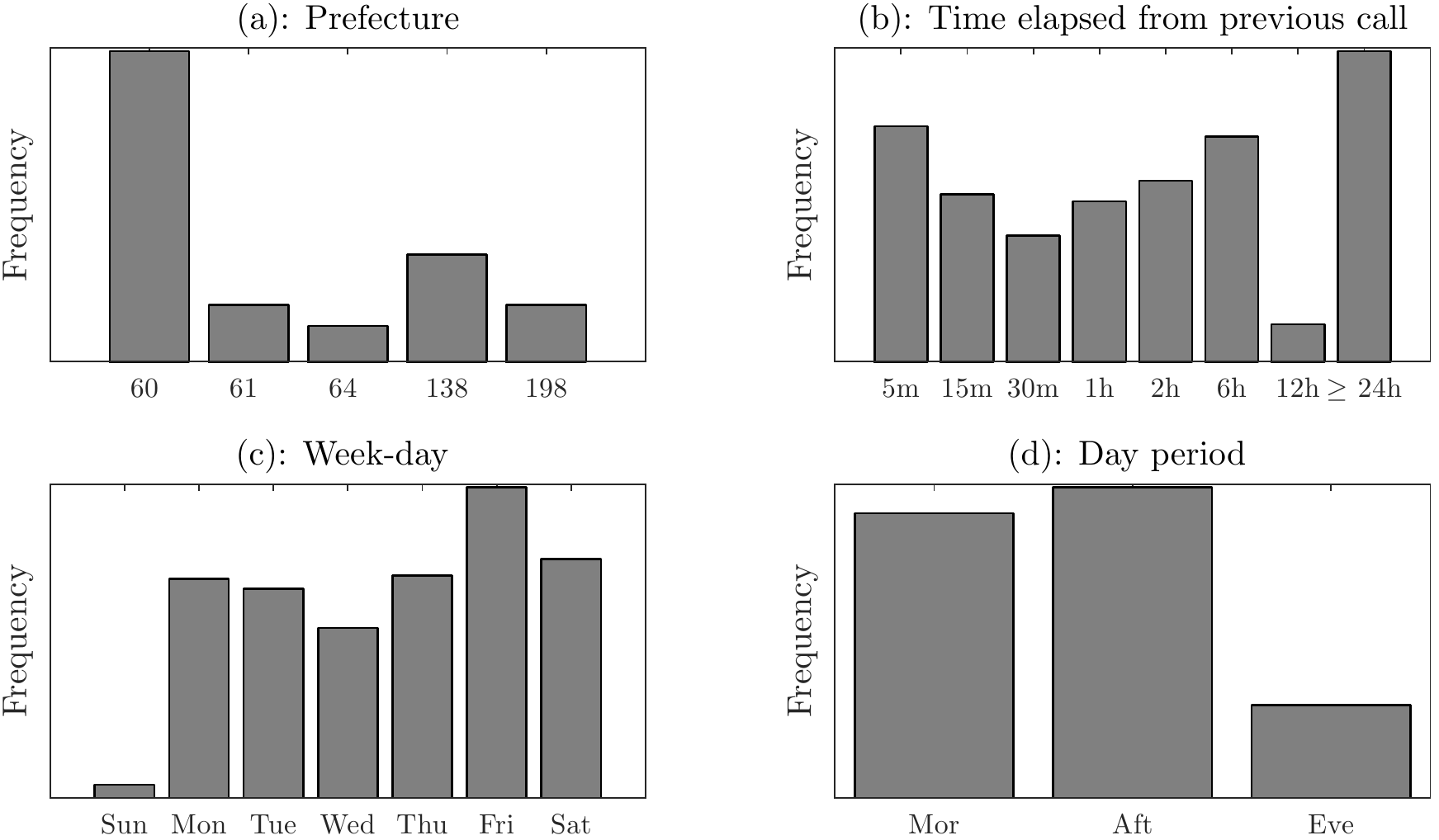}
\caption{These charts represent the distribution of the values in the CDRs of User 4. The chart (a) represents the distribution of the calls among the prefectures visited by the user, the taller the bar in the histogram, the higher the number of calls issued from that prefecture. In (b) we can observe an histogram of the distribution of the calls, which are grouped according to the time elapsed from the previous call. In (c) and (d) the histograms show the total number of calls which are done in the 7 days of the week and in the 3 periods of the day respectively.}
\label{fig:user4graphs}
\end{figure*}

For the sake of conciseness, among all the user that we have processed, we show the results relative to two particular users, characterized by a sufficiently high number of calls, for which we identified meaningful regularities. The identifier of a given user coincides with its order of appearance in the original dataset of 50,000 different customers. The considered users are the 4-th and the 6014-th in the dataset, the number of CDRs for these users are 1453 and 1003, respectively, and their content can be described through the 4 charts that we have previously defined. In Fig. \ref{fig:user4graphs} the values relative to the CDRs of User 4 are depicted using histograms where the value of each bin is normalized with respect to the total number of CDRs. Note that, in order to make the visualization more concise, we do not report the histograms concerning the fields \texttt{work\_day} and \texttt{conn\_time}, which we retained to be the least interesting attributes to be visualized.

From Fig. \ref{fig:user4graphs}(a), it is easy to observe that User 4 performs calls from 5 different prefectures, namely the ones associated with the identifiers 60, 61, 64, 138 and 198 in the dataset, and that most of the calls are issued from prefecture 60, which is likely the one where the person spends most of the time. Note that this spatial information can be easily retrieved from the raw data in the dataset during a pre-processing step. For what concerns the temporal component instead, it is difficult to identify distinct patterns from the original raw data, while they emerge more clearly from the dataset augmented by expliciting the content in the temporal attribute. The distribution of the values contained in such fields is displayed in the 3 remaining histograms. Fig.\ref{fig:user4graphs}(b) depicts an histogram of the distribution of the calls in term of the number of minutes elapsed from the previous call. Every bar represents the number of calls whose time elapsed from the previous call is less than the bin label. The last bin contains all the calls issued after more than 1 day (24 hours) from the previous one. As we can see, the distribution among every bin is well balanced, with the exception of the bin relative to the calls issued after 12 hours, which contains less entries. In \ref{fig:user4graphs}(c) we can observe the distribution of the calls along the days of the week. Even in this case, the number of calls are pretty well distributed with the exception of Sunday, when the activity of the user is very low. Analogously, \ref{fig:user4graphs}(d) represents the distribution of the calls among the 3 periods in which the day is split: a pattern that clearly emerges is that the user calls much more frequently in the morning and in the afternoon, rather than in the evening.

From this first analysis we can draw some initial consideration. For example we can assume that User 4 uses its phone mostly for work calls, which reasonably occur more frequently in the morning and during the work days. In the next sections we perform a more detailed analysis using the two cluster-based algorithms LD-ABCD and PROCLUS.

\subsection{Analysis Results with LD-ABCD}
\begin{figure}[t]
\centering
\includegraphics[width=0.3\textwidth]{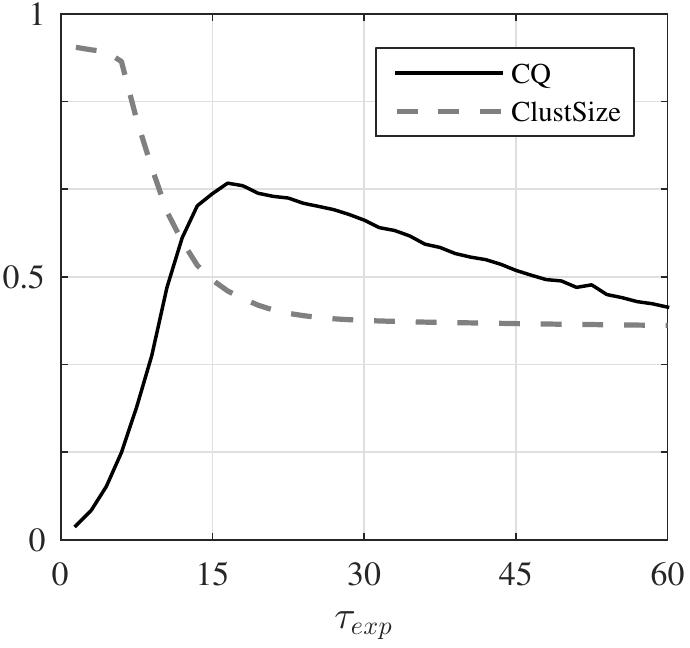}
\caption{The graphic shows how the average CQ and size of the clusters found in the random walk change, as the edges of the considered graph vary according to the different values assumed by $\tau_{\text{exp}}$. The parameter controls the weight of the edges, by modifying the values on their labels. High values of $\tau_{\text{exp}}$ force the walker to visit only very similar nodes, which produce small and compact clusters. On the other hand, a lower value of $\tau_{\text{exp}}$ allows the walker to move on larger sets of nodes, with a consequent discovery of wider clusters, which however are less compact and separated from the remainder of the dataset, and consequently they are characterized by a lower CQ value.}
\label{fig:tauexp}
\end{figure}

The solution returned by LD-ABCD consists in a set of meta-clusters and in the list of associated PCs.
The first step that must be performed before executing the algorithm is the heuristic estimation of the critical parameter $\tau_{\text{exp}}$, according to the procedure described in \cite{bianchi2014agent}. We tried all the values of $\tau_{\text{exp}}$ in the interval $[1,120]$, executing the algorithm 150 different times for each value and we have evaluated the average CQ obtained and average size of the clusters found (see Fig. \ref{fig:tauexp}). 
We noticed that in many dataset of CDRs related to different users, the optimal value of $\tau_{\text{exp}}$ falls in the interval $[15,25]$, for which the average size of the clusters stops to decrease and their average CQ stop to increase.

Concerning the threshold which regulates the minimum allowed CQ value, for low values of $\tau_{\text{CQ}}$ a larger number of clusters is returned and some of them are characterized by a lower quality, in terms of compactness and separability from the remainder of the dataset. This setup is more conservative since less clusters are discarded, but, at the same time, it demands more computational resources because an higher amount of information must be processed in the successive steps of the procedure. For our analysis, we are more concerned about the quality of the result, rather than on computational efficiency. For this reason we set $\tau_{\text{CQ}} = 0.8$, a rather low value, meaning that only the clusters whose CQ is lower than 0.8 are discarded in the searching procedure and are not aggregated to any meta-cluster.

\begin{figure}[!ht]
\centering
\includegraphics[width=0.5\textwidth]{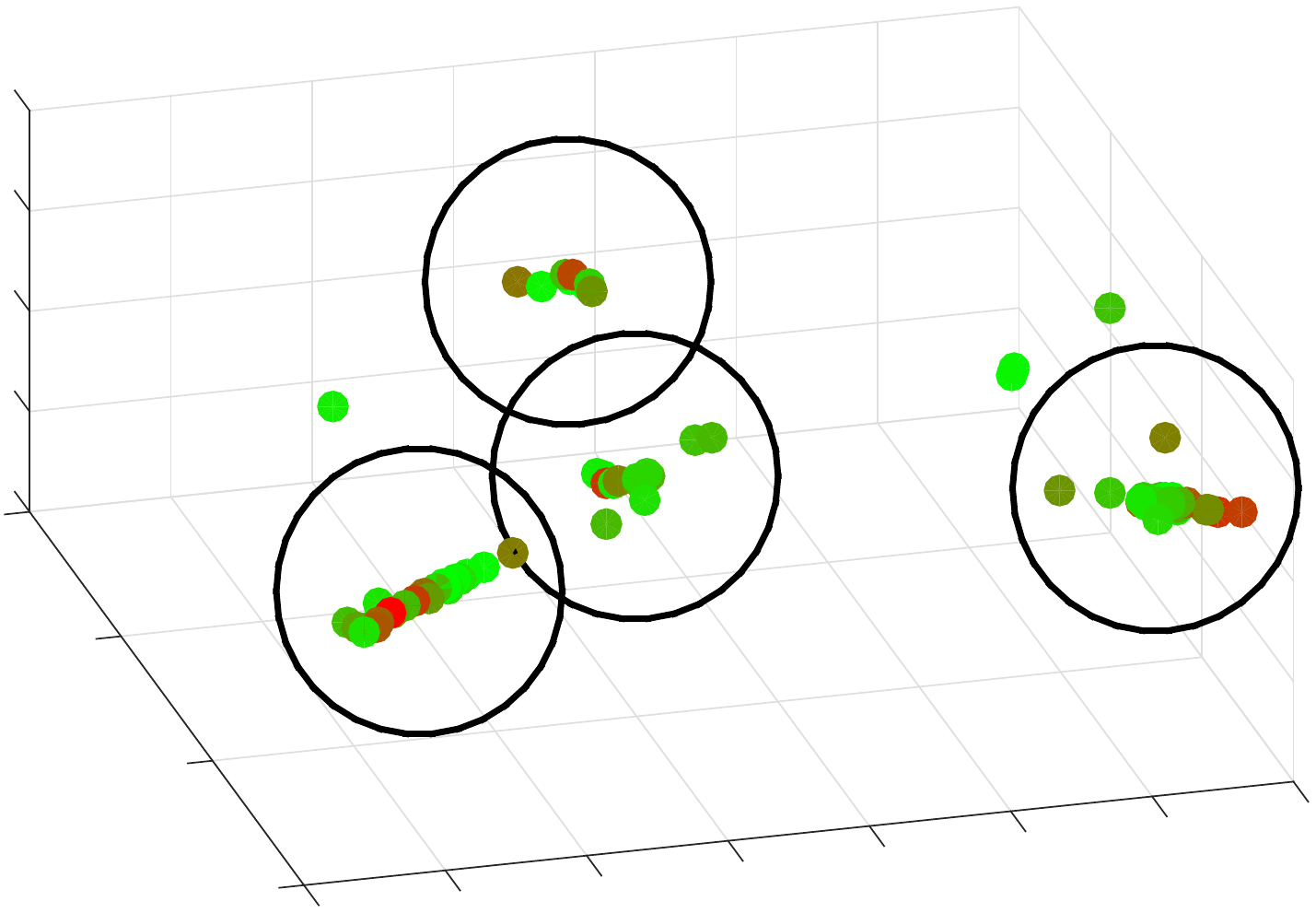}
\caption{Plot of the first 3 principal components resulting from a PCA on $\hat{\mathcal{C}}$, relative to user 4. Each dot represents a meta-cluster and it is possible to identify 4 different dense regions, which are marked with a black circle. The color of a dot represents the CQ value of the meta-clusters: bright red tonalities correspond to high CQ values. }
\label{fig:user4pca}
\end{figure}

The last parameter to be set is the maximum allowed radius of the meta-clusters $\vartheta \in [0,1]$: high values of $\vartheta$ generates less but larger meta-clusters, while for lower values the number of meta-cluster returned is higher, but their dimension is smaller. In our experiments it was useful to retrieve a large number of meta-clusters, in order to identify more clearly the most dense regions of the cluster space, which represent the most recurrent patterns among the CDRs of a given user. Through a trial-and-error approach, we found that setting $\vartheta = 0.2$ generates a number $M$ of meta-clusters which is sufficiently high for the purpose of our analysis. Each meta-cluster $\hat{c}_i$ returned by LD-ABCD is represented with the Boolean vector $\mu_i$ of $N$ elements, where $N$ is the number of CDRs in dataset: if the $j-$th component of the vector $\mu_i(j) = 1$ it means that the $j$-th CDR in the dataset belongs to the cluster, otherwise if $\mu_i(j) = 0$, the $j$-th CDR is not in $\hat{c}_i$. 
In order to visualize the results, we applied a PCA (Principal Component Analysis) to the $M \times N$ matrix $\hat{\mathcal{C}}$ containing all the meta-clusters and we retrieved the first 3 principal components, that is, the representation of $\hat{\mathcal{C}}$ in the first 3 dimensions of the principal component space. In this way, we obtained a $ M \times 3$ matrix where each row corresponds to a meta-cluster that can be visualized in a 3-dimensional space, as depicted in Fig. \ref{fig:user4pca}.
Each meta-cluster in the plot is colored according to its CQ value: a color close to green means that its CQ is low, while red represents the clusters with high CQ values. Note however that all the meta-clusters are formed by clusters whose CQ will be higher than the threshold $\tau_{\text{CQ}}$ and so, even the green dots, represent clusters with a relatively high CQ.

\begin{figure*}
\centering
\subfloat[Meta-cluster $\hat{c}_1$]{\includegraphics[width=0.6\textwidth]{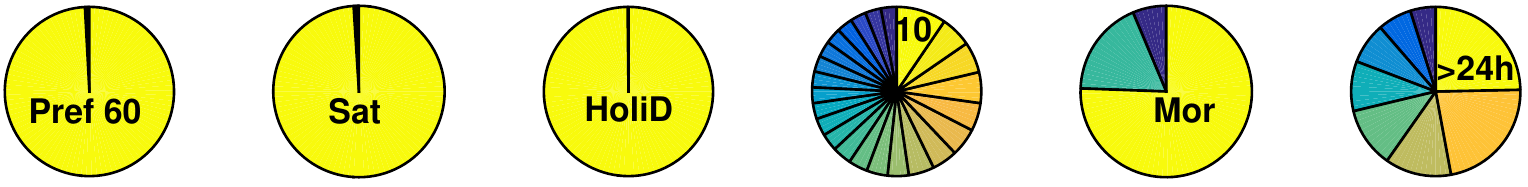}%
\label{fig:LdabcdU4C1}}
\hspace{2em}
\subfloat[Meta-cluster $\hat{c}_2$]{\includegraphics[width=0.6\textwidth]{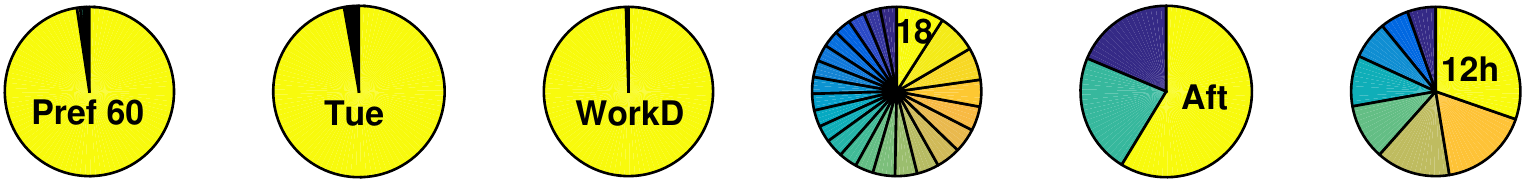}%
\label{fig:LdabcdU4C2}}
\hspace{2em}
\subfloat[Meta-cluster $\hat{c}_3$]{\includegraphics[width=0.6\textwidth]{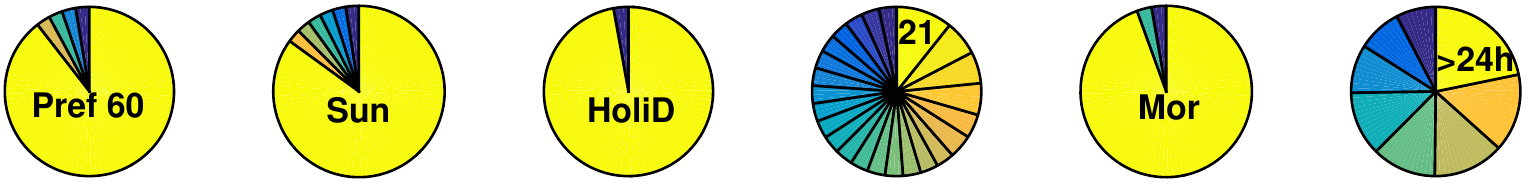}%
\label{fig:LdabcdU4C3}}
\hspace{2em}
\subfloat[Meta-cluster $\hat{c}_4$]{\includegraphics[width=0.6\textwidth]{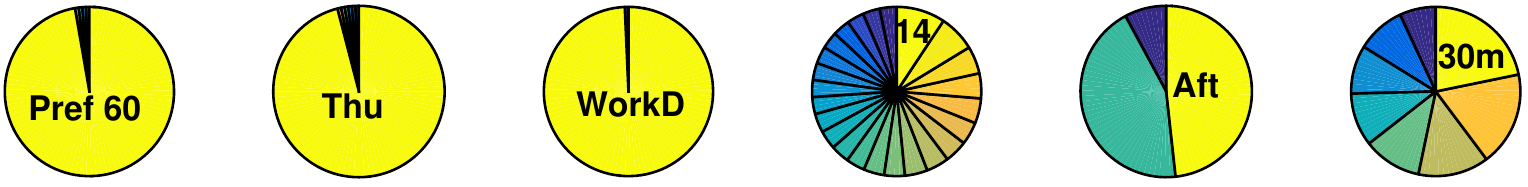}%
\label{fig:LdabcdU4C4}}
\caption{The six pie charts in each row of this figure represent the distribution of the values in the attributes \texttt{subref\_id}, \texttt{week\_day}, \texttt{work\_day},  \texttt{conn\_time}, \texttt{day\_period} and \texttt{prev\_call} in each meta-cluster found in CDRs of User 4. Each slice represents the frequency of the values for that attribute. In each chart we report the name of the most frequent value.}
\label{fig:User4Clust}
\end{figure*}

From the Fig. \ref{fig:user4pca} we can observe that there is a set $\mathcal{R}$ of dense regions, which have been verified to be formed by meta-clusters whose content is similar, i.e. the sets of the elements of the dataset that they represent are strongly overlapped. In this case, we identified 4 dense regions, which are marked in the figure with a black circle. From each region $r \in \mathcal{R}$ we selected the meta-cluster $\hat{c}_r$ with the highest CQ value, (the dot with the brightest red color) and we consider it the representative of the pattern described by the region. 
The list $\mathcal{L}_r$ of PCs associated to $\hat{c}_r$ represents the sets of features according to which the elements contained in $\hat{c}_r$ are similar to each other. Each pair $\langle \hat{c}_r, \mathcal{L}_r \rangle$ represents then a recurrent pattern $r$ in the dataset. In Fig. \ref{fig:User4Clust} we report for each one of the 4 patterns a set of pie charts representing the distribution of the values in \texttt{subref\_id}, \texttt{week\_day}, \texttt{work\_day},  \texttt{conn\_time}, \texttt{day\_period} and \texttt{prev\_call} in the meta-cluster. In the figure, each pie chart represents the values distribution for an attribute. Note that for a more clear interpretation, we labeled only the largest slice, which represent the most frequent value assumed by the attribute.

In the following, we discuss in detail the patterns identified for user 4, according to the content of the meta-clusters.

The first pattern is represented by the meta-cluster $\hat{c}_1$ and the list of PCs $\mathcal{L}_1$, whose content is described in Tab. \ref{tab:L1}. 

\bgroup
\def\arraystretch{1} 
\setlength\tabcolsep{0.2em} 
\begin{center}
\begin{table}[!ht] \footnotesize
\centering
\caption{The table reports the PCs contained in the list $\mathcal{L}_1$ associated to the meta-cluster $\hat{c}_1$ and the average value in each component. Each entry concerns the features \texttt{subref\_id}, \texttt{week\_day}, \texttt{work\_day},  \texttt{conn\_time}, \texttt{day\_period} and \texttt{prev\_call} respectively. A value equal to 1 means that the feature is considered relevant in the cluster, 0 otherwise.}
\begin{tabular}{c|c|c|c|c|c|c|}
\cline{2-7}
& \multicolumn{6}{c|}{\textbf{Values}} \\
\cline{2-7}
& \texttt{subref\_id} & \texttt{week\_day} & \texttt{work\_day} & \texttt{conn\_time} & \texttt{day\_period} & \texttt{prev\_call} \\
\hline
\multicolumn{1}{|c|}{$\text{PC}_1$} & 1 & 1 & 1 & 0 & 1 & 0 \\
\hline
\multicolumn{1}{|c|}{$\text{PC}_2$} & 1 & 1 & 1 & 1 & 0 & 0 \\
\hline
\multicolumn{1}{|c|}{$\text{PC}_3$} & 1 & 1 & 1 & 0 & 1 & 1 \\
\hline
\multicolumn{1}{|c|}{$\text{PC}_4$} & 1 & 1 & 1 & 0 & 0 & 0 \\
\hline
\multicolumn{1}{|c|}{$\text{PC}_5$} & 0 & 1 & 1 & 0 & 1 & 0 \\
\hline
\multicolumn{1}{|c|}{$\text{PC}_6$} & 1 & 0 & 1 & 0 & 0 & 1 \\
\hline
\multicolumn{1}{|c|}{$\text{PC}_7$} & 1 & 1 & 1 & 0 & 0 & 0 \\
\hline
\rowcolor{LightGray}
\multicolumn{1}{|c|}{\textbf{Avg}} & \textbf{0.86} & \textbf{0.86} & \textbf{1} & \textbf{0.14} & \textbf{0.42} & \textbf{0.28} \\
\hline
\end{tabular}
\label{tab:L1}
\end{table}
\end{center}
\egroup
As we can see from Tab. \ref{tab:L1}, the most selected features are \texttt{subref\_id}, \texttt{week\_day} and \texttt{work\_day}. This means that the elements in $\hat{c}_1$ are mostly similar according to these attributes, as confirmed by the charts in Fig. \ref{fig:LdabcdU4C1}. Almost every call comes from the prefecture 60 during Saturday, which is a weekend day. Even if most of the calls are done in the morning and the most frequent time interval from the previous call is 6h, there is an high variance in the values of the attributes as it can be seen from the pie chart, hence the features \texttt{conn\_time}, \texttt{day\_period} and \texttt{prev\_call} are often neglected, by being set to 0 in the related weights of the PCs.

For what concerns the remaining patterns in the CDRs of User 4, the second meta-cluster in Fig. \ref{fig:LdabcdU4C2} highlights the habit of User 4 of performing calls on Tuesday, mostly in the afternoon, from prefecture 60. The third pattern in Fig. \ref{fig:LdabcdU4C3} is related to the calls done in the weekend evening, mostly Saturday from prefecture 60. Finally, the 4-th meta-cluster in Fig. \ref{fig:LdabcdU4C4} represents a recurrent pattern in the telephonic activity of the user, who often calls on Thursday from prefecture 60, mostly in the afternoon.

We can appreciate an important feature of LD-ABCD, by noticing that there are no patterns related, for example, to calls done on Monday, Wednesday or Friday, nor calls done in the evening. This is because there are no strong regularities in the calls involving these days or this period \emph{in conjunction} with other features. In fact, we recall from Sect. \ref{sec:ldabcd_desc} that in LD-ABCD at least 2 elements in the PC must be different from 0, in order to prevent the generation of clusters containing elements similar only with respect to a single feature. For this reason, trivial clusters of CDRs sharing only the day of the week are never considered. 

\begin{figure}[!ht]
\centering
\includegraphics[width=0.5\columnwidth]{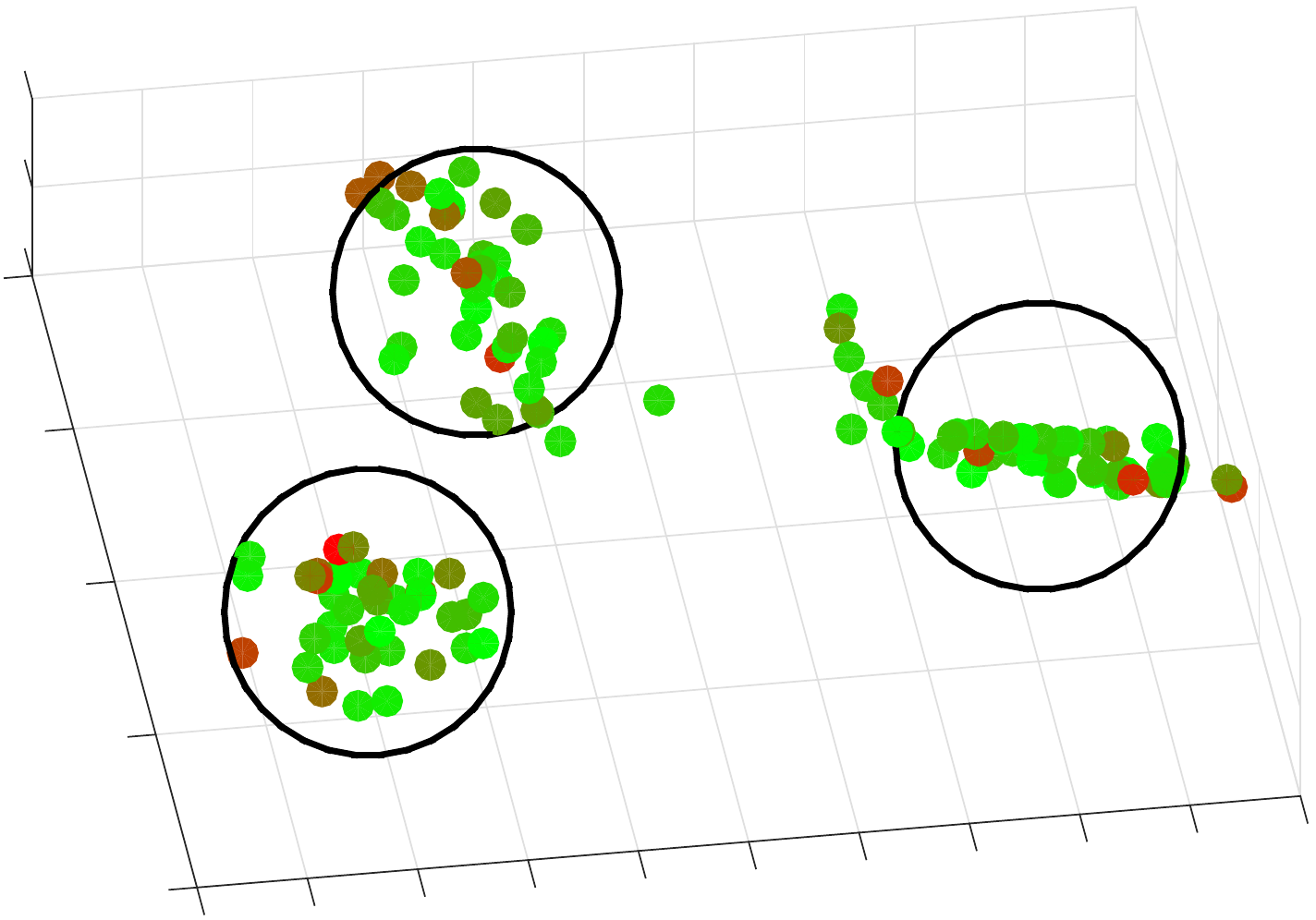}
\caption{Plot of the first 3 principal component scores, resulting from a PCA on the meta-clusters found for user 6014. In this case there are three dense regions, marked with a black circle.}
\label{fig:user6014pca}
\end{figure}

\begin{figure*}
\centering
\subfloat[Meta-cluster $\hat{c}_1$]{\includegraphics[width=0.6\textwidth]{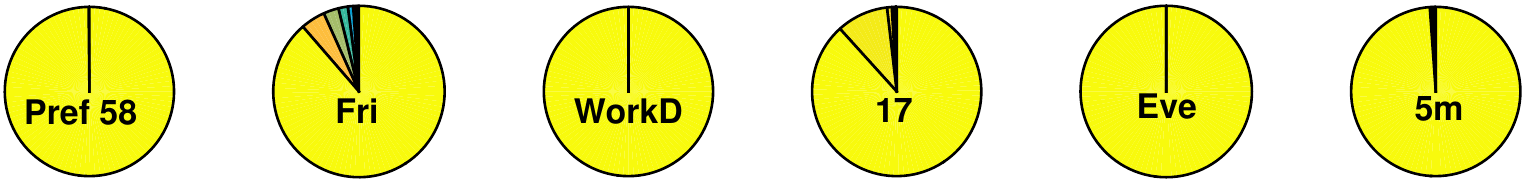}%
\label{fig:LdabcdU6014C1}}
\hspace{2em}
\subfloat[Meta-cluster $\hat{c}_2$]{\includegraphics[width=0.6\textwidth]{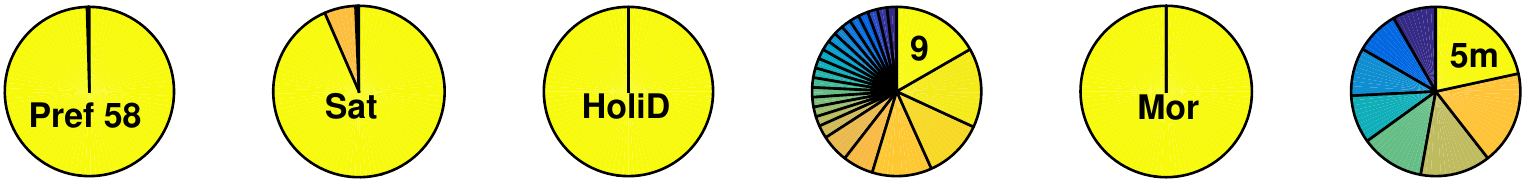}%
\label{fig:LdabcdU6014C2}}
\hspace{2em}
\subfloat[Meta-cluster $\hat{c}_3$]{\includegraphics[width=0.6\textwidth]{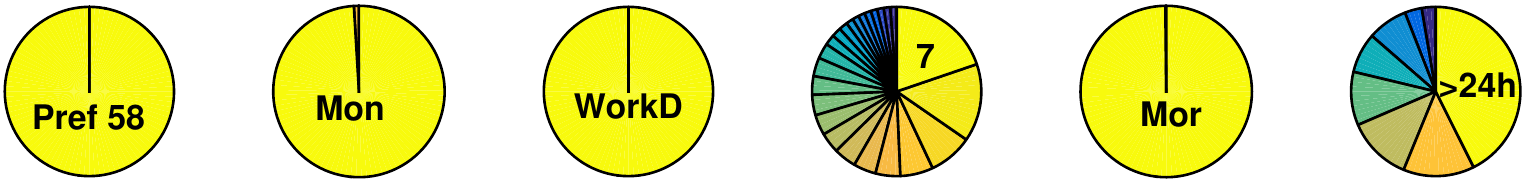}%
\label{fig:LdabcdU6014C3}}
\caption{The set of the 6 pie charts of the meta-clusters that represent the 3 relevant patterns among the CDRs of User 6014.}
\label{fig:User6014Clust}
\end{figure*}

In the following we report another example of analysis, relative to user 6014. Like in the previous case, we plot a visual representation of the meta-clusters using the PCA (see Fig. \ref{fig:user6014pca}) and we identify the most dense regions, which are 3 in this case. From each region we select the meta-clusters with highest CQ value (brightest red color) and we consider them as the representatives of the related 3 recurrent patterns. Again, to describe the content of each meta-cluster and, consequently, the semantic of the related pattern, we use the pie charts in Fig. \ref{fig:User6014Clust}.

In this case, the first meta-cluster in Fig. \ref{fig:LdabcdU6014C1} represents the habit of User 6014 of calling from prefecture 58 in the evening of working days after 5 minutes from the previous call, mostly at 17:00 on Friday. The second pattern in Fig. \ref{fig:LdabcdU6014C2} highlights that the user issues calls in the weekends, during the morning, from prefecture 58, mostly Saturday. Finally, the third meta-cluster in Fig. \ref{fig:LdabcdU6014C3} shows that the user often calls on Monday morning from prefecture 58, mostly after more than 24 hours from the previous call.

We conclude with a consideration on the stability of the results found by LD-ABCD. For each user, we repeated several times the clusters mining procedure and we observed that the number and average quality of the meta-clusters returned in different runs is always the same. This is a valuable result, since the algorithm possess a strong stochastic component, due to the nature of the random walk and the probabilistic selection of new PCs. If the results returned can be repeated it means that the solution is stable and complete, in the sense that most of the recurrent patterns are identified in each run of the algorithm.


\subsection{Analysis Results with PROCLUS}

In the case of study under consideration, which is the identification of relevant and recurrent patterns among the CDRs of a given user, there is not a \textit{ground truth} to which we can refer for evaluating the quality of a solution.
Thereby, in order to appraise the effectiveness of the knowledge discovery system based on the LD-ABCD algorithm, we perform a comparison with the results returned by the procedure based on PROCLUS on the same dataset. In particular, we compare the patterns found by the two algorithms.

As described in Sect. \ref{sec:proclus_desc}, PROCLUS requires the user to specify the number $k$ of clusters to be searched and the average number of dimensions $l$ that must be identified in each cluster. A correct tuning of these parameters requires an \textit{a-priori} knowledge of the problem and of the dataset, which we do not possess. For this reason and for making a more significant comparison between the two algorithms, we set $k$ equal to the number of dense regions found by LD-ABCD in the space of the meta-clusters, after having applied the PCA dimensionality reduction. The parameter $l$ is estimated by considering the average number of features that have been selected by LD-ABCD, i.e. how many features are, in average, set equal to 1 in the PCs associated to each meta-cluster. For what concerns the parameter \texttt{minDev} that controls the number of points in a cluster, we followed the recommendations provided in \cite{Aggarwal:1999:FAP:304181.304188} by the authors of the procedure.

According to the results of LD-ABCD relative to User 4, we set $k=4$ and $l=5$. Like for LD-ABCD, the result of PROCLUS is not deterministic because of the stochastic nature of the initialization procedure. However, it has been proven that in datasets with well defined clusters, each one with a specific set of characterizing dimensions, the results returned in different runs were stable and very similar \cite{yiu2003frequent}. 

We performed a total of 200 different runs on the dataset of the CDRs of User 4, using the same values for the parameters $k$ and $l$, and we analyzed the results obtained. 

\begin{figure}[!ht]
\centering
\includegraphics[width=0.5\columnwidth]{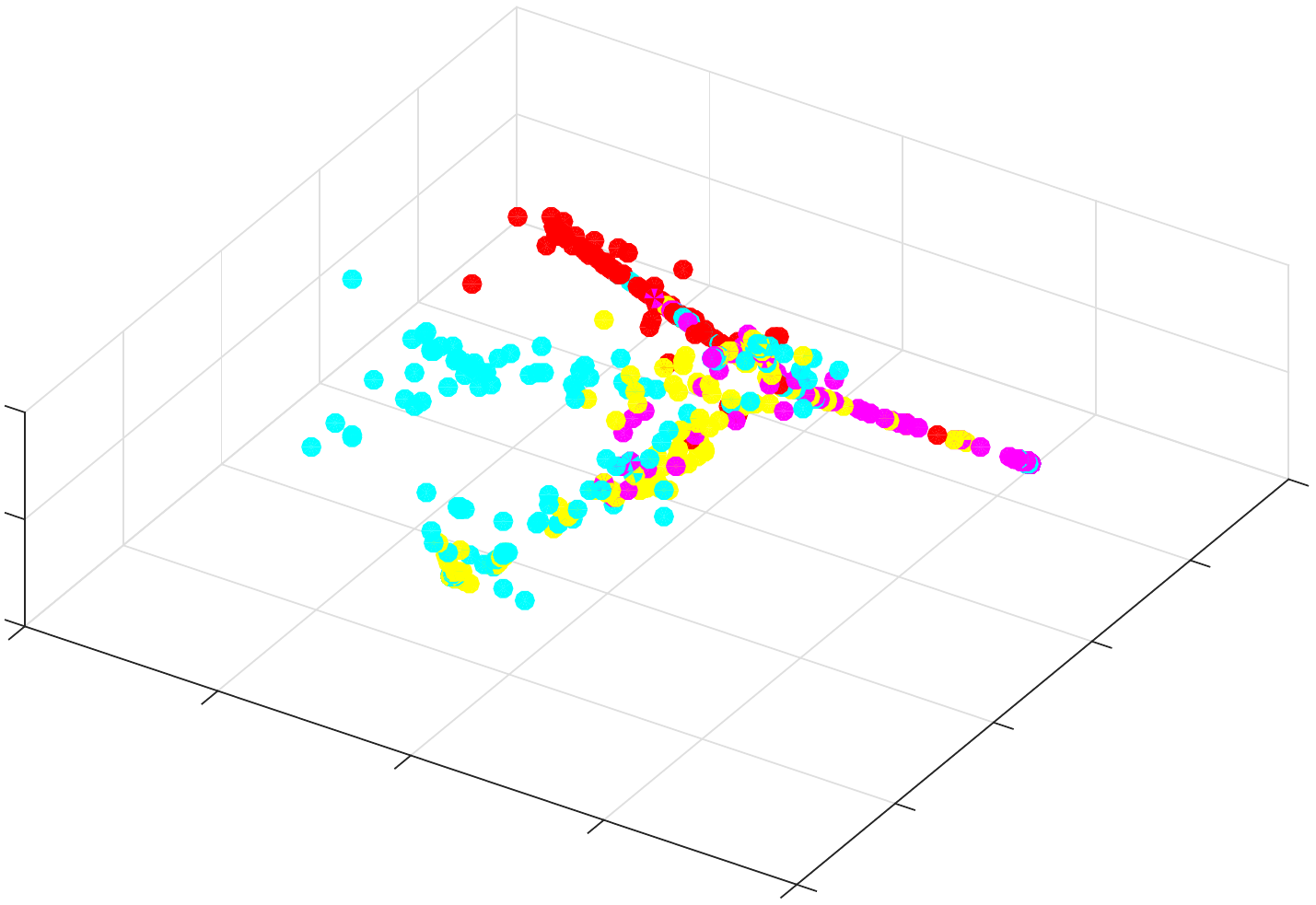}
\caption{The first 3 principal components, resulting from a PCA on the 4 clusters returned by PROCLUS in 200 different runs. For each run, we plotted the 4 clusters found using the same set of colors. Dots of the same colors represent the set of clusters, each one coming from a distinct run, which are the most similar in terms of constituent elements.}
\label{fig:PROCLUSPCA}
\end{figure}

As in the case of LD-ABCD, in Fig. \ref{fig:PROCLUSPCA} we plotted the first 3 principal component scores and we colored differently each one of the 4 clusters returned by a given run, assigning the same color to the most similar clusters among different runs. More specifically, we retrieve the 4 clusters returned by the first run and we assign them 4 different colors: yellow, red, purple and turquoise. Then, in the successive runs we matched the 4 clusters returned with the ones found in the first run. We paired each new cluster with the most similar cluster (in terms of percentage of shared elements) from the first run using a Best Match First heuristic \cite{4403133, BianchiGranular2015} and we assigned the same 4 colors to the new clusters, according to this match.
As we can see from Fig. \ref{fig:PROCLUSPCA}, even if clusters of the same color (which are the most similar with respect to the matching) are mostly located in the same areas, we cannot distinguish 4 distinct dense groups and, especially in the middle, the clusters are strongly overlapped and mixed. This means that in different runs PROCLUS is not able to identify the same set of clusters, like LD-ABCD does.

\begin{figure}[!ht]
\centering
\includegraphics[width=0.6\columnwidth]{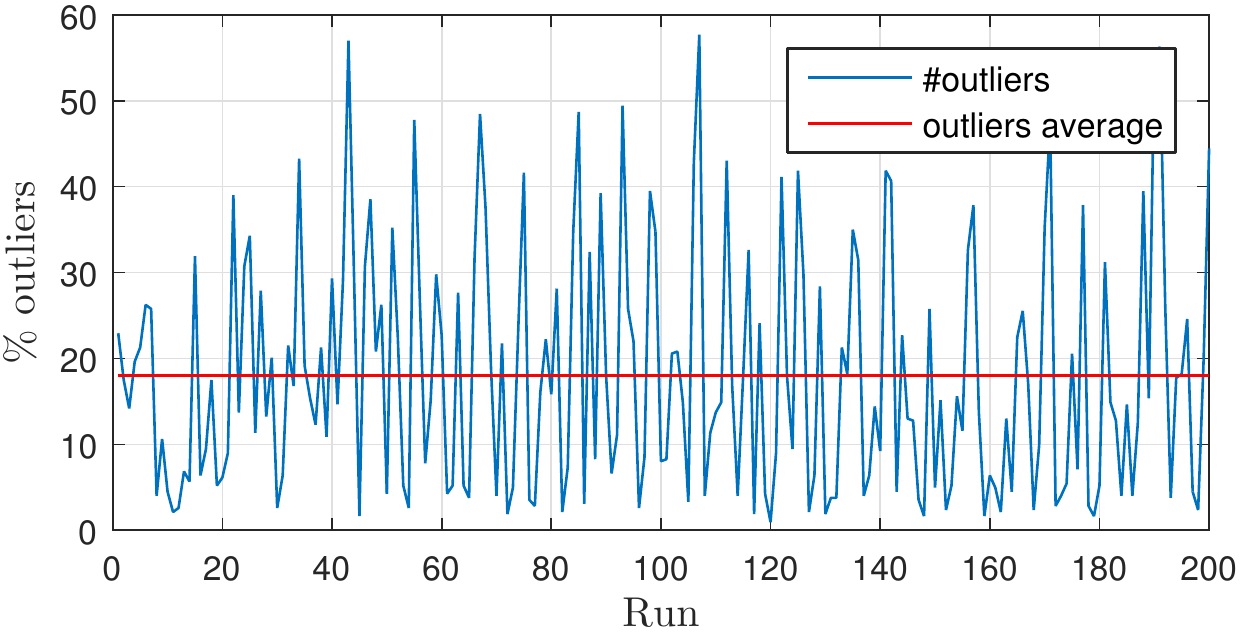}
\caption{Number of outliers, expressed as a percentage of the total number of elements in the dataset, identified by PROCLUS in 200 different runs; in red is plotted the average value. As we can observe the number of outliers varies considerably from run to run, and in some cases it reaches very high values, up to 60 \% of the elements in the dataset.}
\label{fig:PROCLUSOutliers}
\end{figure}

Another symptom of the instability of the solution returned is the high variance of the number of outliers that the algorithm finds in different runs. In Fig. \ref{fig:PROCLUSOutliers} we plotted the percentage of outliers found in each run and the average value. We can observe that there is a very high variance in the number of elements that the algorithm recognizes as outliers. If the composition of the outliers set changes significantly in each run, it means that the structure of the remaining clusters varies as well. Thus, there is not a stable, unique solution identified by the algorithm. 

Several clusters found by PROCLUS results to be accurate, in terms of the similarity of the attributes of the elements in each cluster. However such clusters are generally small and they are returned in runs where an high percentage of elements are classified as outliers. In larger cluster instead, the values in the attributes of the contained elements are generally more heterogeneous, even if the PROCLUS  selects such attributes as relevant dimensions in the cluster. 
\begin{figure*}[!ht]
\centering
\subfloat[Meta-cluster A]{\includegraphics[width=0.6\textwidth]{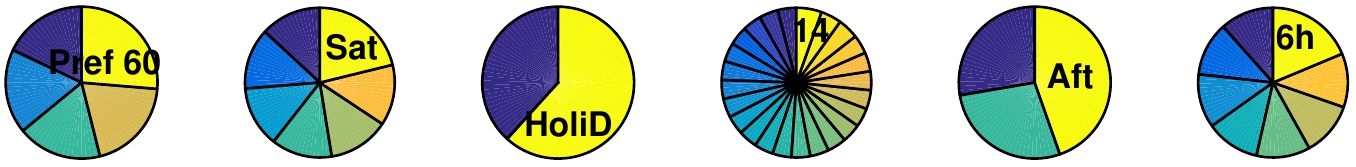}%
\label{fig:ProclusU4C1}}
\hspace{2em}
\subfloat[Meta-cluster B]{\includegraphics[width=0.6\textwidth]{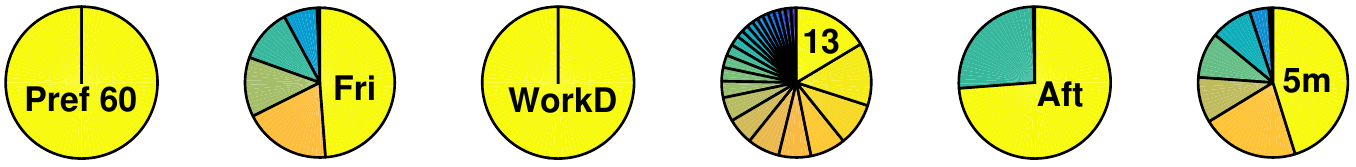}%
\label{fig:ProclusU4C2}}
\caption{Two different clusters identified by PROCLUS. In the first row there are the pie charts relative to a cluster A of large size and, as we can observe, the attributes on the different values are very heterogeneous among the elements in the cluster. In the second row a cluster B is represented with graphics that show a higher accuracy on the attributes, but the size of this cluster is significantly smaller.}
\label{fig:PROCLUSC4}
\end{figure*}
As an example, in the first row of Fig. \ref{fig:PROCLUSC4} we depict the pie charts relative to a large cluster A, whose dimension is comparable in size with the clusters found by LD-ABCD. However, as we can see from the charts, the values of different attributes among the elements in cluster A are very heterogeneous. On the second row instead, the pie charts represent a cluster B which contains elements with more homogeneous values on the relevant attributes, but the dimension of cluster B is much smaller, since it contains approximately 7\% of the total CDRs and it comes from a run with a very high percentage of outliers, which is 23\% of the entire dataset.

\section{Conclusions and Future Works}
\label{sec:conclusion}

In this paper we applied our recently developed knowledge discovery algorithm LD-ABCD, as the core engine for a data mining analysis on CDRs. The objective of this work is to build profiles of the users, which can be employed by telecommunication companies for monitoring and understanding the behaviors of their subscribers.
A suitable characterization of the customer base allows to use with greater effectiveness the information resources of a telecom operator, to develop marketing strategies and for tailoring telephonic plans which better suit the needing of the users. Furthermore, a series of applications can be designed for monitoring the activity of a user, focused on the identification of anomalous and suspicious behaviors, which are the ones which differs from the usual subscribers' patterns of habits and that could compromise his privacy or security.

We analyzed the CDRs from the dataset of the D4D challenge, which contains only two type of values, concerning the time and the geographical position from where the calls were issued. We showed how additional features can be derived from these original attributes and how regularities can be extracted from a dataset which apparently contain a very limited quantity of information.

LD-ABCD is an agent-based algorithm that identify regularities and recurrent patterns among data. When applied to a dataset of CDRs, it can be used to identify habits in telephonic activity of the users, in order to create their "digital-fingerprints". One of most important features of the LD-ABCD, in our applicative context, is the possibility of identifying multiple parameter configurations, which highlight the characteristics of patterns within a cluster that are considered to be discriminative. Such configurations represent the key for interpreting and characterizing semantically the regularities found in the dataset. Another important advantage of LD-ABCD, with respect to other approaches to cluster analysis, is that the number $k$ of clusters to be identified is not an input parameter of the algorithm, but it is automatically identified during the discovery procedure. This is particularly useful when there are no information on the number of possible clusters to be identified in the data, like in our case of study.

We compared the results of the knowledge discovery system based on LD-ABCD with an alternative implementation based on PROCLUS, the well-known subspace clustering algorithm capable of identifying clusters in a subset of the original feature-space. Both LD-ABCD and PROCLUS share the important characteristic of being able to identify a local metric for each retrieved cluster. 
We discussed the results of the analysis considering the CDRs of a specific user, applying the knowledge discovery systems based on LD-ABCD and PROCLUS and we showed how LD-ABCD is capable of identifying a set of patterns which can be semantically characterized. 
The result returned by LD-ABCD demonstrated to be stable and reliable, even when the nature of the data is completely unknown and when the presence of well-defined clusters is not clear. This is also a consequence of the low sensitivity of LD-ABCD to different settings of the configuration parameters ($\tau_{\text{exp}}$, $\tau_{\text{CQ}}$ and $\vartheta$), compared to the case of PROCLUS, where different choices of $k$ and $l$ could significantly modify the results. Additionally, as we discussed in Sect. \ref{sect:experiments}, the most critical parameter $\tau_{\text{exp}}$ can be easily tuned, using an heuristic procedure, which demonstrated to be effective in different contexts \cite{bianchi2014agent}.
 
On the other side, even if LD-ABCD can be easily implemented on a distributed computing network, due to the multi-agent architecture, the main disadvantage of the procedure is the high computational resources required, in terms of both space and time. For this reason, when the nature of the problem faced is simpler or when prior information on the problem are provided, many other alternatives can be considered for implementing the core engine of a data mining procedure. The difficulty of the considered problem has been confirmed by the results returned by PROCLUS, which was not able to identify a stable set of clusters among the data. In fact, from the experiments we observed that PROCLUS, in different execution of the clustering procedure, was not able to identify the same set of clusters, showing an high variance and instability in the results returned. 

The results obtained confirm the effectiveness of the LD-ABCD and they encourage further applications.
In a future work, we firstly plan to process the CDRs of every user in the dataset, searching for common patterns and regularities in order to group together users characterized by a similar profile. Then, in a second step, with the identified clusters of users we aim to define specific classes, which can be analyzed and used for describing some general, common behaviors that allow a better understanding of the habits of the customer base of a telecommunication company. Important information can be assessed concerning the geographical location of the user and a study can be conducted on how the habits change in different areas. Finally, we plan to consider additional datasets relative to a telecommunication network, which are characterized by a large amount of entries and a higher number of features (for example, those included in the Telecom Italia BigData Challenges \cite{telecomBigData1, telecomBigData2}).

\bibliographystyle{abbrv}
\bibliography{Bibliography}

\begin{thebibliography}{10}

\bibitem{telecomBigData1}
{ Telecom Italia}.
\newblock {Telecom Italia} big data challenge 2014, 2014.
\newblock Accessed: 2015-11-16.

\bibitem{Aggarwal:1999:FAP:304181.304188}
C.~C. Aggarwal, J.~L. Wolf, P.~S. Yu, C.~Procopiuc, and J.~S. Park.
\newblock Fast algorithms for projected clustering.
\newblock {\em SIGMOD Rec.}, 28:61--72, June 1999.

\bibitem{agrawal1998automatic}
R.~Agrawal, J.~Gehrke, D.~Gunopulos, and P.~Raghavan.
\newblock {\em Automatic subspace clustering of high dimensional data for data
  mining applications}, volume~27.
\newblock ACM, 1998.

\bibitem{ahas2010using}
R.~Ahas, S.~Silm, O.~J{\"a}rv, E.~Saluveer, and M.~Tiru.
\newblock Using mobile positioning data to model locations meaningful to users
  of mobile phones.
\newblock {\em Journal of Urban Technology}, 17(1):3--27, 2010.

\bibitem{Bereta20131213}
M.~Bereta, W.~Pedrycz, and M.~Reformat.
\newblock {Local descriptors and similarity measures for frontal face
  recognition: A comparative analysis}.
\newblock {\em Journal of Visual Communication and Image Representation},
  24(8):1213--1231, 2013.

\bibitem{berlingerio2013allaboard}
M.~Berlingerio, F.~Calabrese, G.~{Di Lorenzo}, R.~Nair, F.~Pinelli, and M.~L.
  Sbodio.
\newblock {AllAboard: a system for exploring urban mobility and optimizing
  public transport using cellphone data}.
\newblock In {\em {Machine Learning and Knowledge Discovery in Databases}},
  pages 663--666. Springer, 2013.

\bibitem{berry2011computational}
D.~M. Berry.
\newblock The computational turn: Thinking about the digital humanities.
\newblock {\em Culture Machine}, 12(0), 2011.

\bibitem{bianchi2014agent}
F.~Bianchi, E.~Maiorino, L.~Livi, A.~Rizzi, and A.~Sadeghian.
\newblock An agent-based algorithm exploiting multiple local dissimilarities
  for clusters mining and knowledge discovery.
\newblock {\em Soft Computing}, pages 1--23, 2015.

\bibitem{7286732}
F.~M. Bianchi, E.~De~Santis, A.~Rizzi, and A.~Sadeghian.
\newblock Short-term electric load forecasting using echo state networks and
  {PCA} decomposition.
\newblock {\em IEEE Access}, 3:1931--1943, Oct. 2015.

\bibitem{bianchi2013matching}
F.~M. Bianchi, L.~Livi, and A.~Rizzi.
\newblock {Matching of time-varying labeled graphs}.
\newblock In {\em {Neural Networks (IJCNN), The 2013 International Joint
  Conference on}}, pages 1--8. IEEE, 2013.

\bibitem{bianchi2014granular}
F.~M. Bianchi, L.~Livi, A.~Rizzi, and A.~Sadeghian.
\newblock {A Granular Computing approach to the design of optimized graph
  classification systems}.
\newblock {\em Soft Computing}, 18(2):393--412, 2014.

\bibitem{BianchiGranular2015}
F.~M. Bianchi, S.~Scardapane, A.~Rizzi, A.~Uncini, and A.~Sadeghian.
\newblock Granular computing techniques for classication and semantic
  characterization of structured data.
\newblock {\em Cognitive Computation}, Dec 2015.

\bibitem{bianchi2015prediction}
F.~M. Bianchi, S.~Scardapane, A.~Uncini, A.~Rizzi, and A.~Sadeghian.
\newblock Prediction of telephone calls load using {E}cho {S}tate {N}etwork
  with exogenous variables.
\newblock {\em Neural Networks}, 71:204--213, 2015.

\bibitem{DBLP:journals/corr/abs-1210-0137}
V.~D. Blondel, M.~Esch, C.~Chan, F.~Cl{\'{e}}rot, P.~Deville, E.~Huens,
  F.~Morlot, Z.~Smoreda, and C.~Ziemlicki.
\newblock Data for development: the {D4D} challenge on mobile phone data.
\newblock {\em CoRR}, abs/1210.0137, 2012.

\bibitem{calabrese2011real}
F.~Calabrese, M.~Colonna, P.~Lovisolo, D.~Parata, and C.~Ratti.
\newblock Real-time urban monitoring using cell phones: A case study in rome.
\newblock {\em Intelligent Transportation Systems, IEEE Transactions on},
  12(1):141--151, 2011.

\bibitem{candia2008uncovering}
J.~Candia, M.~C. Gonz{\'a}lez, P.~Wang, T.~Schoenharl, G.~Madey, and A.-L.
  Barab{\'a}si.
\newblock Uncovering individual and collective human dynamics from mobile phone
  records.
\newblock {\em Journal of Physics A: Mathematical and Theoretical},
  41(22):224015, 2008.

\bibitem{chang2012boosting}
C.-C. Chang.
\newblock {A boosting approach for supervised Mahalanobis distance metric
  learning}.
\newblock {\em Pattern Recognition}, 45(2):844--862, 2012.

\bibitem{4403133}
G.~Del~Vescovo and A.~Rizzi.
\newblock Automatic classification of graphs by symbolic histograms.
\newblock In {\em Granular Computing, 2007. GRC 2007. IEEE International
  Conference on}, pages 410--410, Nov 2007.

\bibitem{demirli2000higher}
K.~Demirli and P.~Muthukumaran.
\newblock Higher order fuzzy system identification using subtractive
  clustering.
\newblock {\em Journal of Intelligent \& Fuzzy Systems: Applications in
  Engineering and Technology}, 9(3, 4):129--158, 2000.

\bibitem{duong2010constructing}
T.~H. Duong, N.~T. Nguyen, and G.~S. Jo.
\newblock Constructing and mining a semantic-based academic social network.
\newblock {\em Journal of Intelligent \& Fuzzy Systems: Applications in
  Engineering and Technology}, 21(3):197--207, 2010.

\bibitem{eagle2009inferring}
N.~Eagle, A.~S. Pentland, and D.~Lazer.
\newblock Inferring friendship network structure by using mobile phone data.
\newblock {\em Proceedings of the National Academy of Sciences},
  106(36):15274--15278, 2009.

\bibitem{fayyad2002information}
U.~M. Fayyad, A.~Wierse, and G.~G. Grinstein.
\newblock {\em Information visualization in data mining and knowledge
  discovery}.
\newblock Morgan Kaufmann, 2002.

\bibitem{furletti2013analysis}
B.~Furletti, L.~Gabrielli, C.~Renso, and S.~Rinzivillo.
\newblock Analysis of gsm calls data for understanding user mobility behavior.
\newblock In {\em Big Data, 2013 IEEE International Conference on}, pages
  550--555. IEEE, 2013.

\bibitem{citeulike:11530535}
H.~Gao, J.~Tang, and H.~Liu.
\newblock {Mobile Location Prediction in Spatio-Temporal Context}.
\newblock {\em the Procedings of Mobile Data Challenge by Nokia Workshop at the
  Tenth International Conference on Pervasive Computing}, June 2012.

\bibitem{gidofalvi2012and}
G.~Gid{\'o}falvi and F.~Dong.
\newblock When and where next: individual mobility prediction.
\newblock In {\em Proceedings of the First ACM SIGSPATIAL International
  Workshop on Mobile Geographic Information Systems}, pages 57--64. ACM, 2012.

\bibitem{gonen2011multiple}
M.~G{\"o}nen and E.~Alpayd{\i}n.
\newblock {Multiple kernel learning algorithms}.
\newblock {\em The Journal of Machine Learning Research}, 12:2211--2268, 2011.

\bibitem{grindrod2013infering}
P.~Grindrod.
\newblock Infering behavior-based lifestyle categorizations based on mobile
  phone usage data, Mar.~15 2013.
\newblock US Patent App. 13/841,852.

\bibitem{hu2013we}
D.~Hu, F.~Sun, L.~Tu, and B.~Huang.
\newblock We know what you are--a user classification based on mobile data.
\newblock In {\em Green Computing and Communications (GreenCom), 2013 IEEE and
  Internet of Things (iThings/CPSCom), IEEE International Conference on and
  IEEE Cyber, Physical and Social Computing}, pages 1282--1289. IEEE, 2013.

\bibitem{jain1988algorithms}
A.~K. Jain and R.~C. Dubes.
\newblock {\em Algorithms for clustering data}.
\newblock Prentice-Hall, Inc., 1988.

\bibitem{japkowicz1999concept}
N.~Japkowicz.
\newblock {\em Concept-learning in the absence of counter-examples: an
  autoassociation-based approach to classification}.
\newblock PhD thesis, Rutgers, The State University of New Jersey, 1999.

\bibitem{kannan2004clusterings}
R.~Kannan, S.~Vempala, and A.~Vetta.
\newblock {On clusterings: Good, bad and spectral}.
\newblock {\em Journal of the ACM (JACM)}, 51(3):497--515, 2004.

\bibitem{kim_duin__multiplediss__2009}
S.-W. Kim and R.~P.~W. Duin.
\newblock {A Combine-Correct-Combine Scheme for Optimizing Dissimilarity-Based
  Classifiers}.
\newblock In E.~Bayro-Corrochano and J.-O. Eklundh, editors, {\em {Progress in
  Pattern Recognition, Image Analysis, Computer Vision, and Applications}},
  volume 5856 of {\em {LNCS}}, pages 425--432. Springer Berlin Heidelberg,
  2009.

\bibitem{komorowski1997principles}
J.~Komorowski and J.~Zytkow.
\newblock {\em Principles of data mining and knowledge discovery}.
\newblock Springer, 1997.

\bibitem{lazer2009life}
D.~Lazer, A.~S. Pentland, L.~Adamic, S.~Aral, A.~L. Barabasi, D.~Brewer,
  N.~Christakis, N.~Contractor, J.~Fowler, M.~Gutmann, et~al.
\newblock Life in the network: the coming age of computational social science.
\newblock {\em Science (New York, NY)}, 323(5915):721, 2009.

\bibitem{li2002dpf}
B.~Li, E.~Chang, and C.-T. Wu.
\newblock {DPF-a perceptual distance function for image retrieval}.
\newblock In {\em {Proceedings of the 2002 International Conference on Image
  Processing.}}, volume~2, pages II--597. IEEE, 2002.

\bibitem{lima2013exploiting}
A.~Lima, M.~{De Domenico}, V.~Pejovic, and M.~Musolesi.
\newblock {Exploiting cellular data for disease containment and information
  campaigns strategies in country-wide epidemics}.
\newblock {\em arXiv preprint arXiv:1306.4534}, 2013.

\bibitem{olsson2004fault}
E.~Olsson, P.~Funk, and N.~Xiong.
\newblock Fault diagnosis in industry using sensor readings and case-based
  reasoning.
\newblock {\em Journal of Intelligent \& Fuzzy Systems}, 15(1):41--46, 2004.

\bibitem{ozer2014predicting}
M.~{\"O}zer.
\newblock {\em PREDICTING THE LOCATION AND TIME OF MOBILE PHONE USERS BY USING
  SEQUENTIAL PATTERN MINING TECHNIQUES}.
\newblock PhD thesis, Middle East Technical University, 2014.

\bibitem{parsons2004subspace}
L.~Parsons, E.~Haque, and H.~Liu.
\newblock Subspace clustering for high dimensional data: a review.
\newblock {\em ACM SIGKDD Explorations Newsletter}, 6(1):90--105, 2004.

\bibitem{6399461}
W.~Pedrycz.
\newblock {Proximity-Based Clustering: A Search for Structural Consistency in
  Data With Semantic Blocks of Features}.
\newblock {\em IEEE Transactions on Fuzzy Systems}, 21(5):978--982, 2013.

\bibitem{Queiroz20133383}
S.~Queiroz, F.~d. A.~T. {de Carvalho}, and Y.~Lechevallier.
\newblock {Nonlinear multicriteria clustering based on multiple dissimilarity
  matrices}.
\newblock {\em Pattern Recognition}, 46(12):3383--3394, 2013.

\bibitem{schlaich2010generating}
J.~Schlaich, T.~Otterst{\"a}tter, and M.~Friedrich.
\newblock Generating trajectories from mobile phone data.
\newblock In {\em Proceedings of the 89th Annual Meeting Compendium of Papers,
  Transportation Research Board of the National Academies}, 2010.

\bibitem{6588959}
C.~Shen, J.~Kim, F.~Liu, L.~Wang, and A.~{van den Hengel}.
\newblock {Efficient Dual Approach to Distance Metric Learning}.
\newblock {\em IEEE Transactions on Neural Networks and Learning Systems},
  25(2):394--406, Feb 2014.

\bibitem{smith2013ubiquitous}
C.~Smith, A.~Mashhadi, and L.~Capra.
\newblock Ubiquitous sensing for mapping poverty in developing countries.
\newblock {\em Paper submitted to the Orange D4D Challenge}, 2013.

\bibitem{song2010limits}
C.~Song, Z.~Qu, N.~Blumm, and A.-L. Barab{\'a}si.
\newblock Limits of predictability in human mobility.
\newblock {\em Science}, 327(5968):1018--1021, 2010.

\bibitem{Spinosa:2007:OCA:1244002.1244107}
E.~J. Spinosa, A.~P. de~Leon F.~de Carvalho, and J.~a. Gama.
\newblock Olindda: A cluster-based approach for detecting novelty and concept
  drift in data streams.
\newblock In {\em Proceedings of the 2007 ACM Symposium on Applied Computing},
  SAC '07, pages 448--452, New York, NY, USA, 2007. ACM.

\bibitem{taormina2015neural}
R.~Taormina and K.~Chau.
\newblock Neural network river forecasting with multi-objective fully informed
  particle swarm optimization.
\newblock {\em J. Hydroinform}, 17:99--113, 2015.

\bibitem{telecomBigData2}
{Telecom Italia}.
\newblock {Telecom Italia} big data challenge 2015, 2015.
\newblock Accessed: 2015-11-16.

\bibitem{StefvandenElzenD4D}
S.~van~den Elzen, J.~Blaas, J.~J. van Wijk, and R.~Spousta.
\newblock {Exploration and Analysis of Massive Mobile Phone Data: A Layered
  Visual Analytics approach}.
\newblock to apper, May 2013.

\bibitem{wang2015improving}
W.-c. Wang, K.-w. Chau, D.-m. Xu, and X.-Y. Chen.
\newblock Improving forecasting accuracy of annual runoff time series using
  arima based on eemd decomposition.
\newblock {\em Water Resources Management}, 29(8):2655--2675, 2015.

\bibitem{yager1994generation}
R.~R. Yager and D.~P. Filev.
\newblock Generation of fuzzy rules by mountain clustering.
\newblock {\em Journal of Intelligent \& Fuzzy Systems: Applications in
  Engineering and Technology}, 2(3):209--219, 1994.

\bibitem{4674367}
L.~Yang, R.~Jin, L.~Mummert, R.~Sukthankar, A.~Goode, B.~Zheng, S.~C.~H. Hoi,
  and M.~Satyanarayanan.
\newblock {A Boosting Framework for Visuality-Preserving Distance Metric
  Learning and Its Application to Medical Image Retrieval}.
\newblock {\em IEEE Transactions on Pattern Analysis and Machine Intelligence},
  32(1):30--44, Jan 2010.

\bibitem{yin2012semi}
X.~Yin, T.~Shu, and Q.~Huang.
\newblock {Semi-supervised fuzzy clustering with metric learning and entropy
  regularization}.
\newblock {\em Knowledge-Based Systems}, 35:304--311, 2012.

\bibitem{yiu2003frequent}
M.~L. Yiu and N.~Mamoulis.
\newblock Frequent-pattern based iterative projected clustering.
\newblock In {\em Data Mining, 2003. ICDM 2003. Third IEEE International
  Conference on}, pages 689--692. IEEE, 2003.

\bibitem{zhang2012semi}
H.~Zhang, J.~Yu, M.~Wang, and Y.~Liu.
\newblock {Semi-supervised distance metric learning based on local linear
  regression for data clustering}.
\newblock {\em Neurocomputing}, 93:100--105, 2012.

\bibitem{zhang2009multilayer}
J.~Zhang, K.-W. Chau, et~al.
\newblock Multilayer ensemble pruning via novel multi-sub-swarm particle swarm
  optimization.
\newblock {\em J. UCS}, 15(4):840--858, 2009.

\bibitem{zhang2009dimension}
S.~Zhang and K.-W. Chau.
\newblock Dimension reduction using semi-supervised locally linear embedding
  for plant leaf classification.
\newblock In {\em Emerging Intelligent Computing Technology and Applications},
  pages 948--955. Springer, 2009.

\end{thebibliography}
\end{document}